\documentclass[a4paper,fleqn]{cas-dc}
\usepackage{acro}
\acsetup{single}

\DeclareAcronym{OFAC}{
  short = OFAC,
  long  = Treasury's Office of Foreign Assets Control,
}
\newcommand{\OFAC}{\ac{OFAC}\xspace}

\DeclareAcronym{CFTC}{
  short = CFTC,
  long  = Commodity Futures Trading Commission,
}
\newcommand{\CFTC}{\ac{CFTC}\xspace}

\DeclareAcronym{SEC}{
  short = SEC,
  long  = Securities and Exchange Commission,
}
\newcommand{\SEC}{\ac{SEC}\xspace}

\DeclareAcronym{FSB}{
  short = FSB,
  long  = Financial Stability Board,
}
\newcommand{\FSB}{\ac{FSB}\xspace}

\DeclareAcronym{FSMA}{
  short = FSMA,
  long  = Financial Services and Markets Act 2000,
}
\newcommand{\FSMA}{\ac{FSMA}\xspace}

\DeclareAcronym{MAR}{
  short = MAR,
  long  = Market Abuse Regulation,
}
\newcommand{\MAR}{\ac{MAR}\xspace}

\DeclareAcronym{MiCA}{
  short = MiCA,
  long  = Markets in Crypto-Assets,
}
\newcommand{\MiCA}{\ac{MiCA}\xspace}

\DeclareAcronym{FinCEN}{
  short = FinCEN,
  long  = Financial Crimes Enforcement Network,
}
\newcommand{\FinCEN}{\ac{FinCEN}\xspace}

\DeclareAcronym{SEA}{
  short = SEA,
  long  = Securities Exchange Act of 1934,
}
\newcommand{\SEA}{\ac{SEA}\xspace}

\DeclareAcronym{CEA}{
  short = CEA,
  long  = Commodity Exchange Act,
}
\newcommand{\CEA}{\ac{CEA}\xspace}

\DeclareAcronym{HMT}{
  short = HMT,
  long  = HM Treasury,
}
\newcommand{\HMT}{\ac{HMT}\xspace}

\DeclareAcronym{AML}{
  short = AML,
  long  = Anti-Money Laundering,
}
\newcommand{\AML}{\ac{AML}\xspace}

\DeclareAcronym{KYC}{
  short = KYC,
  long  = Know Your Customer,
}
\newcommand{\KYC}{\ac{KYC}\xspace}

\DeclareAcronym{CTF}{
  short = CTF,
  long  = Counter-Terrorist Financing,
}
\newcommand{\CTF}{\ac{CTF}\xspace}

\DeclareAcronym{FCA}{
  short = FCA,
  long  = Financial Conduct Authority,
}
\newcommand{\FCA}{\ac{FCA}\xspace}

\DeclareAcronym{FIEA}{
  short = FIEA,
  long  = Financial Instruments and Exchange Act,
}
\newcommand{\FIEA}{\ac{FIEA}\xspace}

\DeclareAcronym{PSA}{
  short = PSA,
  long  = Payment Service Act,
}
\newcommand{\PSA}{\ac{PSA}\xspace}

\DeclareAcronym{FSA}{
  short = FSA,
  long  = Financial Services Agency,
}
\newcommand{\FSA}{\ac{FSA}\xspace}

\DeclareAcronym{ROI}{
  short = ROI,
  long  = Return On Investment,
}
\newcommand{\ROI}{\ac{ROI}\xspace}

\DeclareAcronym{DeFi}{
  short = DeFi,
  long  = Decentralized Finance,
}
\newcommand{\DeFi}{\ac{DeFi}\xspace}

\DeclareAcronym{TradFi}{
  short = TradFi,
  long  = Traditional Finance,
}
\newcommand{\TradFi}{\ac{TradFi}\xspace}

\DeclareAcronym{PGA}{
  short = PGA,
  long  = Priority Gas Auction,
}
\newcommand{\PGA}{\ac{PGA}\xspace}
\newcommand{\PGAs}{\acp{PGA}\xspace}

\DeclareAcronym{PoW}{
  short = PoW,
  long  = Proof-of-Work,
}
\newcommand{\PoW}{\ac{PoW}\xspace}

\DeclareAcronym{PoS}{
  short = PoS,
  long  = Proof-of-Stake,
}

\DeclareAcronym{SGA}{
  short = SGA,
  long  = Sealed-Bid Auction,
}
\newcommand{\SGA}{\ac{SGA}\xspace}
\newcommand{\SGAs}{\acp{SGA}\xspace}

\DeclareAcronym{JIT}{
  short = JIT,
  long  = Just-in-Time,
}
\newcommand{\JIT}{\ac{JIT}\xspace}

\DeclareAcronym{DEX}{
  short = DEX,
  long  = Decentralized Exchange,
}
\newcommand{\DEX}{\ac{DEX}\xspace}
\newcommand{\DEXs}{\acp{DEX}\xspace}

\DeclareAcronym{CEX}{
  short = CEX,
  long  = Centralized Exchange,
}
\newcommand{\CEX}{\ac{CEX}\xspace}
\newcommand{\CEXs}{\acp{CEX}\xspace}

\DeclareAcronym{EOA}{
  short = EOA,
  long  =  Externally-Owned Account,
}

\DeclareAcronym{NFT}{
  short = NFT,
  long  = Non-fungible Token,
}

\DeclareAcronym{LP}{
  short = LP,
  long  = Liquidity Provider,
}
\newcommand{\LP}{\ac{LP}\xspace}
\newcommand{\LPs}{\acp{LP}\xspace}

\DeclareAcronym{P2P}{
  short = P2P,
  long  = Peer-to-Peer,
}
\newcommand{\PtP}{\ac{P2P}\xspace}

\DeclareAcronym{TVL}{
  short = TVL,
  long  = Total Value Locked,
}

\DeclareAcronym{CFMM}{
  short = CFMM,
  long  = Constant Function Market Maker,
}

\DeclareAcronym{CPMM}{
  short = CPMM,
  long  = Constant Product Market Maker,
}

\DeclareAcronym{DApp}{
  short = DApp,
  long  = Decentralized Application,
}
\newcommand{\DApp}{\ac{DApp}\xspace}
\newcommand{\DApps}{\acp{DApp}\xspace}

\DeclareAcronym{POF}{
  short = POF,
  long  = Private Order Flow,
}
\newcommand{\POF}{\ac{POF}\xspace}
\newcommand{\POFs}{\acp{POF}\xspace}

\DeclareAcronym{CeFi}{
  short = CeFi,
  long  = Centralized Finance,
}
\newcommand{\CeFi}{\ac{CeFi}\xspace}

\DeclareAcronym{MEV}{
  short = MEV,
  long  = Maximal Extractable Value,
}
\newcommand{\MEV}{\ac{MEV}\xspace}

\DeclareAcronym{EV}{
  short = EV,
  long  = Extractable Value,
}

\DeclareAcronym{BEV}{
  short = BEV,
  long  = Blockchain Extractable Value,
}

\DeclareAcronym{AMM}{
  short = AMM,
  long  = Automated Market Maker,
}

\DeclareAcronym{FaaS}{
  short = FaaS,
  long  = Front-running as a Service,
}
\newcommand{\FaaS}{\ac{FaaS}\xspace}
\newcommand{\FaaSs}{\acp{FaaS}\xspace}

\DeclareAcronym{ZKP}{
  short = ZKP,
  long  = Zero-knowledge Proof,
}

\DeclareAcronym{TC}{
  short = TC,
  long  = Tornado.Cash,
}
\newcommand{\TC}{\ac{TC}\xspace}

\DeclareAcronym{DAO}{
  short = DAO,
  long  = Decentralized Autonomous Organization,
}

\newcommand{\DAOs}{\acp{DAO}\xspace}

\newcommand{\ETH}{\ensuremath{\xspace\texttt{ETH}}\xspace}

\newcommand{\USDC}{\ensuremath{\xspace\texttt{USDC}}\xspace}

\newcommand{\BTC}{\ensuremath{\xspace\texttt{BTC}}\xspace}
\newcommand{\USD}{\ensuremath{\xspace\texttt{USD}}\xspace}
\newcommand{\USDT}{\ensuremath{\xspace\texttt{USDT}}\xspace}
\newcommand{\fUSDC}{\ensuremath{\xspace\texttt{fUSDC}}\xspace}
\newcommand{\WBTC}{\ensuremath{\xspace\texttt{WBTC}}\xspace}
\newcommand{\renBTC}{\ensuremath{\xspace\texttt{renBTC}}\xspace}
\newcommand{\KYL}{\ensuremath{\xspace\texttt{KYL}}\xspace}
\newcommand{\WETH}{\ensuremath{\xspace\texttt{WETH}}\xspace}

\newcommand{\XRP}{\ensuremath{\xspace\texttt{XRP}}\xspace}

\newcommand{\tx}{\mathsf{tx}\xspace}

\newcommand{\numJITs}{\empirical{$36{,}671$}\xspace}

\newcommand{\avgProfitETH}{\empirical{$0.204$}\xspace}

\newcommand{\firstBot}{\href{https://etherscan.io/address/0xa57bd00134b2850b2a1c55860c9e9ea100fdd6cf}{0xa57...6CF}\xspace}

\newcommand{\numSandwich}{\empirical{$208{,}149$}\xspace}

\newcommand{\swapRatioSdw}{\empirical{$6$}\xspace}
\newcommand{\capitalInvSdw}{\empirical{$8.37$}\xspace}

\newcommand{\totProfitSandwich}{\empirical{$12{,}242$}\xspace}
\newcommand{\avgProfitSandwich}{\empirical{$0.059$}\xspace}

\newcommand{\avgROISandwich}{\empirical{$1.629\%$}\xspace}
\newcommand{\avgROI}{\empirical{$0.007\%$}\xspace}

\newcommand{\bribeRatioSandwich}{\empirical{$11.5\%$}\xspace}

\newcommand{\firstBotProfitPCT}{\empirical{$92\%$}\xspace}
\newcommand{\swapRatio}{\empirical{$269$}\xspace}

\newcommand{\dateStart}{\empirical{Jun~$1$,~$2021$}\xspace}

\newcommand{\dateEnd}{\empirical{Jan~$31$,~$2023$}\xspace}

\usepackage{nomencl}
\usepackage{etoolbox}
\makeatletter
\appto\@floatboxreset{%
  \ifx\@captype\andy@table
    \mdseries
  \fi
}
\def\andy@table{table}
\makeatother

\usepackage{bm}
\usepackage{textcomp}
\usepackage{color}
\usepackage{setspace}
\usepackage[authoryear]{natbib}
\usepackage{algpseudocode}

\usepackage{mathtools}
\usepackage{longtable}
\usepackage{tabularx}
\usepackage{makecell}
\usepackage{pifont}
\usepackage{xspace}

\usepackage{nicefrac}
\usepackage{siunitx}
\usepackage{array,framed}
\usepackage{flushend}
\usepackage{graphicx}
\usepackage{appendix}

\usepackage{microtype}
\usepackage{setspace}
\usepackage{booktabs}

\usepackage{tikz}
\usetikzlibrary{fit}
\newcommand*\ec[1][1ex]{\tikz\draw (0,0) circle (#1);} 
\newcommand*\hc[1][1ex]{%
  \begin{tikzpicture}
  \draw[fill] (0,0)-- (90:#1) arc (90:270:#1) -- cycle ;
  \draw (0,0) circle (#1);
  \end{tikzpicture}}
\newcommand*\fc[1][1ex]{\tikz\fill (0,0) circle (#1);} 

\newcommand\addvmargin[1]{%
  \node[fit=(current bounding box),inner ysep=#1,inner xsep=0]{};
}

\newcommand{\boldline}[1]{%
  \begin{tikzpicture}
    \expandafter\draw\expandafter[#1] (0,0) -- (0.24,0);
  \addvmargin{0.5mm}
  \end{tikzpicture}%
}

\usepackage{listings}
\usepackage{xspace}

\usepackage{bigstrut}
\usepackage{filecontents}
\usepackage{booktabs}

\usepackage{bigstrut}
\usepackage{filecontents}
\usepackage{booktabs}

\usepackage{subcaption}
\usepackage{amsfonts}
\usepackage{amsthm}
\usepackage{multirow}
\usepackage[ruled,vlined]{algorithm2e}
\usepackage{algpseudocode}
\usepackage{graphicx}
\usepackage{enumitem}
\usepackage{numprint}
\usepackage[normalem]{ulem}
\usepackage{tcolorbox}
\usepackage{lipsum}

\usepackage{xcolor}
\usepackage{colortbl}
\usepackage{tabularray}

\theoremstyle{definition}
\newtheorem{definition}{Definition}[section]

\newcommand{\empirical}[1]{{\color{black}#1}}

\definecolor{gainsboro}{rgb}{0.86, 0.86, 0.86}
\newcommand{\greycol}[1]{\cellcolor{gainsboro}{#1}}

\definecolor{whitegray}{rgb}{0.95, 0.95, 0.95}

\usepackage[n,landau,notions,ff,mm]{cryptocode}
\definecolor{gamechangecolor}{gray}{0.74}

\newcommand*\bcircled[1]{\tikz[baseline=(char.base)]{
    \node[fill=black,text=white, shape=circle,draw=black,inner sep=.6pt] (char) {#1};}}

\begin{document}
\let\WriteBookmarks\relax
\def\floatpagepagefraction{1}
\def\textpagefraction{.001}

\shorttitle{Market Misconduct in Decentralized Finance}

\shortauthors{Xihan Xiong et~al.}

\title [mode = title]{Market Misconduct in Decentralized Finance (DeFi): Analysis, Regulatory Challenges and Policy Implications}

\author[1]{Xihan Xiong}

\affiliation[1]{organization={Department of Computing, Imperial College London},
    addressline={Exhibition Rd, South Kensington}, 
    city={London},
    postcode={SW7 2BX}, 
    country={United Kingdom}}

\author[1]{Zhipeng Wang}

\author[2]{Tianxiang Cui}
\cormark[1]


\ead{tianxiang.cui@nottingham.edu.cn}



\affiliation[2]{organization={School of Compute Science, University of Nottingham Ningbo China},
    addressline={199 Taikang E Rd}, 
    city={Ningbo, Zhejiang},
    postcode={315104}, 
    country={China}} 

\author[1]{William Knottenbelt}

\author[1]{Michael Huth}

\begin{abstract}
Technological advancement drives financial innovation, reshaping the traditional finance (TradFi) system and redefining user-market interactions. The rise of blockchain technology and Decentralized Finance (DeFi) stand as prime examples of such progress. While DeFi has introduced opportunities, it has also exposed the ecosystem to new forms of market misconduct, which remain inadequately regulated. In this paper, we explore the potential of blockchain technology to facilitate market misconduct within DeFi, compare misconduct in the TradFi and DeFi markets, identify emerging forms of DeFi market misconduct, and discuss the regulatory challenges and policy implications. We hope this study will provide policymakers with insights on bringing DeFi into the regulatory perimeter. 
\end{abstract}


\maketitle


\section{Introduction} \label{sec:intro}

Technological innovation has emerged as a disruptive force, revolutionizing various aspects of the financial industry. The recent development has been remarkable, particularly in the areas of blockchain and \DeFi~(\citealp{allen2020blockchain,ciarli2021digital,hojckova2020entrepreneurial}). Blockchain technology, the foundational technology behind cryptocurrencies, enables secure, transparent, and decentralized transactions. It eliminates the need for financial intermediaries by utilizing a distributed network to verify and record transactions. The decentralization characteristic of blockchain has unlocked the opportunity for financial innovations, particularly in the \DeFi space, a growing ecosystem of financial products and services built on top of permissionless blockchains. Numerous innovative products have emerged within the \DeFi landscape. \DeFi innovations are transformative, encompassing stablecoins that operate in a decentralized manner while pegged to fiat currencies, lending platforms enabling peer-to-peer borrowing without intermediaries, and \DEXs allowing traders to exchange assets directly from their wallets. \DeFi has attracted substantial adoption in recent years. The total value locked in \DeFi platforms reached $87$b~\USD in Aug $2024$\footnote{\href{https://www.statista.com/statistics/1272181/defi-tvl-in-multiple-blockchains/}{Statista, TVL across multiple DeFi blockchains}.}.

Yet, integrating technology into finance comes with its challenges. Technological advancements, while facilitating innovation, have also paved the way for new forms of financial market misconduct. \DeFi is a case in point. 
For instance, the bZx incident\footnote{\href{https://peckshield.medium.com/bzx-hack-ii-full-disclosure-with-detailed-profit-analysis-8126eecc1360}{bZx Hack II Full Disclosure}.}, which occurred in February 2020, involved an attacker using a flash loan to manipulate asset prices and extract profits, demonstrating how this \DeFi innovation can become a tool for exploitation.
One of the key impacts of this incident was the proliferation of new types of market misconduct in \DeFi. The incident served as a wake-up call for developers and regulators. It indicates that malicious actors can exploit vulnerabilities in the system to engage in various forms of misconduct. Consequently, this raised important policy implications, emphasizing the need for regulatory frameworks to address the risks associated with \DeFi.

Indeed, in response to the emerging risks in \DeFi, the UK \HMT took proactive steps by releasing a consultation on the regulatory regime for crypto-assets\footnote{\href{https://www.gov.uk/government/consultations/future-financial-services-regulatory-regime-for-cryptoassets}{Future financial services regulatory regime for cryptoassets}.}. The \HMT intends to achieve consistent regulatory outcomes for comparable activities in \DeFi and \TradFi, despite adopting different regulatory tools and timelines. Furthermore, the U.S. Department of the Treasury has issued a report\footnote{\href{https://home.treasury.gov/system/files/136/DeFi-Risk-Full-Review.pdf}{Illicit Finance Risk Assessment of Decentralized Finance}.} that examines the way in which malicious actors exploit the vulnerabilities associated with \DeFi services, highlighting the importance of addressing potential challenges related to \AML and \CTF measures for \DeFi.

Even with the early regulatory response, a notable regulatory gap persists. While \TradFi is subject to extensive regulatory frameworks, \DeFi still lacks specific regulatory measures. In \TradFi, regulations such as insider dealing prohibitions, market manipulation rules, and disclosure requirements are firmly established to protect market integrity. Regulatory bodies such as the \SEC in the U.S. and the \FCA in the UK diligently oversee and implement these rules. However, in the emerging field of \DeFi, the lack of specific regulations exposes users to various forms of market misconduct. In addition, although academic research has thoroughly investigated market misconduct in \TradFi (e.g., \citealp{cumming2011exchange,cumming2012high,Cumming2015financial,khodabandehlou2022market,putnicnvs2020overview,alexander2022corruption,dalko2018how,Dalko2020High}), DeFi market misconduct remains inadequately studied.

Bridging the academic and regulatory gap is crucial to mitigate the risks associated with \DeFi. This study aims to systematically analyze \DeFi market misconduct and evaluate its implications for regulatory frameworks. We intend to tackle the following research questions: 

\vspace{+1mm}
\noindent\textbf{RQ1: How could the misuse or exploitation of blockchain technology exacerbate market misconduct in DeFi?} We examine the characteristics of blockchain decentralization, pseudonymity, transparency, and atomicity, and how their combination with \DeFi's composability may give rise to novel forms of market misconduct. 
In particular, we investigate how mixing services, flash loans, \FaaSs, and transaction auctions can be exploited to facilitate market misconduct within the \DeFi ecosystem.

\vspace{+1mm}
\noindent\textbf{RQ2: What forms of market misconduct in \DeFi are comparable to those observed in \TradFi?} We first provide an overview of market misconduct in \TradFi, covering various forms such as market manipulation and insider dealing. Next, we present the definition of \DeFi market misconduct that encompasses market efficiency, fairness, and \DeFi security. Furthermore, we provide a comprehensive taxonomy of \DeFi market misconduct. Based on the definition and taxonomy, we compare the market misconduct in \TradFi and \DeFi, highlighting the key areas where they differ. 

\vspace{+1mm}
\noindent\textbf{RQ3: What novel forms of misconduct have emerged in the \DeFi market, facilitated by its unique characteristics and decentralized nature?} We investigate the emergence of novel misconduct forms in the \DeFi landscape. Additionally, we present a case study demonstrating how multiple manipulation techniques can be combined to exploit vulnerabilities in the \DeFi ecosystem, emphasizing the interplay between different forms of misconduct.

\vspace{+1mm}
\noindent\textbf{RQ4: What are the regulatory challenges and policy implications of \DeFi market misconduct?} 
We first outline the regulatory challenges and risks. Then we analyze initial regulatory responses and identify areas where existing regulatory frameworks may need enhancement. Finally, we propose potential strategies to integrate DeFi within the regulatory perimeter and identify future research avenues. 

\smallskip The remainder of the paper is organized as follows. 
 Section~\ref{sec:background} introduces the background information. We discuss the concerns on DeFi market misconduct in Section~\ref{sec: concerns}.  Sections~\ref{sec: misconduct-tradfi} and~\ref{sec: misconduct-defi} compare the misconduct in \TradFi and \DeFi. Section~\ref{sec: policy} presents the policy implications and regulatory challenges. Section~\ref{sec: discussion} and Section~\ref{sec:limitation} discuss the impact and limitations. Section~\ref{sec: conclusion} concludes this paper.

\section{The New Financial Marketplace} \label{sec:background}

\subsection{Blockchain Primer}

\subsubsection{Blockchain}
A blockchain functions as a distributed ledger maintained by nodes within a \PtP network. Transactions are organized into blocks, which are linked sequentially through cryptographic hash connections, ensuring the integrity and immutability of recorded data.
This design provides a transparent, tamper-resistant, decentralized system to record and verify transactions. Specifically, every participant on the blockchain network is allowed to issue and broadcast transactions. 
On a \PoW blockchain (e.g., Bitcoin), block creators, known as miners, solve computational \PoW puzzles to verify transactions and add new blocks. In contrast, on a Proof-of-Stake (PoS) blockchain (e.g., Ethereum), block creators are called validators. They stake \ETH into the deposit contract, which serves as collateral that can be foreclosed if the validator engages in dishonest behavior. The validators verify transactions and vote on the outcomes to append new blocks to the chain.

\subsubsection{Smart Contract}

A smart contract contains a set of predefined conditions, automatically executed once the specified conditions are met. Smart contracts eliminate the need for third-party intermediaries to enforce contract terms, enhancing operational efficiency. Once deployed on the blockchain, smart contracts are self-executing, meaning they automatically execute their predefined functions without human intervention. They are also self-enforcing, as the code and logic embedded in the contract ensure that parties involved in the agreement adhere to the terms without relying on trust alone. All contract terms, conditions, and actions are recorded on the blockchain and can be audited and verified by all participants. 

On smart contract-based blockchains, users can trigger the smart contract execution by issuing transactions. If the transaction fails because a specific condition cannot be met, the entire transaction will not be executed, and all accounts will revert to the prior state. Even if the transaction is reverted, the issuer will still be charged for the gas fees.

\subsubsection{\PtP Network}

The \PtP network serves as a foundational element of blockchain technology. This network facilitates direct communication and interactions among nodes. In a blockchain \PtP network, each node maintains a complete copy of the blockchain ledger. Information sharing is facilitated through specialized network protocols designed for blockchain operations, such as Bitcoin and Ethereum protocols. Transactions are automatically propagated among nodes, while miners or validators monitor the network to receive pending transactions and validate their inclusion in the blockchain.

\subsubsection{Blockchain Transaction}
\begin{figure}[tbh]
\centering
\includegraphics[width=\columnwidth]{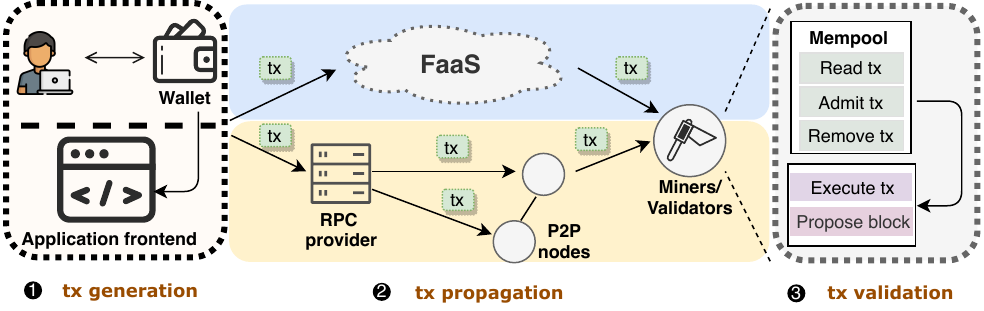}
\caption{The blockchain transaction life cycle includes transaction generation, propagation, and validation. Figure adapted from (adapted from~\cite{wang2023blockchain}).}
\label{fig:tx_flow}
\end{figure}

The life cycle of a blockchain transaction consists of three stages: generation, propagation, and validation~(see Figure~\ref{fig:tx_flow}). During transaction generation, a user's wallet creates a transaction ($\tx$) by signing it with a private key. The wallet then transmits $\tx$ to an RPC provider or node, which propagates it across the \PtP network. Transactions are initially distributed across either a public \PtP network or a \FaaS before being verified by miners or validators~(\citealp{kim2018measuring}).
Once a transaction is validated, its content is recorded by nodes that participated in its propagation. \emph{Pending transactions} are those that have been submitted but not yet confirmed in a block. When initiated, transactions are broadcast to the network and temporarily stored in the \emph{mempool}, a queue for unconfirmed transactions. Pending transactions arise because blockchain networks have limited capacity to process transactions per block. Miners or validators select transactions from the mempool based on factors such as fees, priority, and the network's consensus mechanism. Transactions with higher fees or priority are typically processed first.

\subsection{The DeFi landscape}
\subsubsection{DeFi Reference Frame}\label{sec:defi_ref_frame}

We introduce a DeFi-tailored reference frame proposed by \cite{zhou2022sok}, consisting of five layers. The network layer (NEN) encompasses the underlying infrastructure and protocols that enable the communication and interaction between participants in a blockchain network. The consensus layer (CON) ensures agreement and consistency across the distributed network of nodes. The smart contract layer (SC) enables the execution of self-executing and tamper-proof contracts. The \DeFi protocol layer (PRO) refers to the collection of \DApps that provide various financial services and products in a decentralized and permissionless manner. Auxiliary services (AUX) encompass entities to ensure the efficiency of \DeFi but do not fall within the scope of the aforementioned system layers. 

\subsubsection{DeFi}
Smart contracts facilitate the creation of financial products and services on top of permissionless blockchains, known as \DeFi. It transforms the \CeFi system by creating open and transparent financial applications and protocols such as \DEX, lending, and stablecoins. \DeFi has gained significant popularity in recent years, driven by its potential to revolutionize traditional financial systems and offer innovative solutions to users. 

\subsubsection{TradFi, CeFi and DeFi}
\TradFi, \CeFi and \DeFi are conceptually different. Comprehending the differences between them helps users effectively engage with these systems. A comparison of their key characteristics is provided in Table~\ref{tab:comparison}.

\begin{table*}[t]
\centering
 \resizebox{0.98\linewidth}{!}{ 
\begin{tabular}{l|lll}
\hline
\greycol{Property} & \greycol{TradFi} & \greycol{CeFi} & \greycol{DeFi} \\ \hline

\makecell[l]{Financial Products \\\& Services} & \makecell[l]{Traditional financial products and services} & \makecell[l]{Tokenized traditional financial products \\or cryptocurrencies trading on CEXs} & \makecell[l]{Innovative and decentralized financial \\products and services trading on DEXs} \\ \hline

\makecell[l]{Control \\\& Governance} & \makecell[l]{Centralized control and governance by \\financial institutions and regulators} & \makecell[l]{Centralized control and governance by \\financial institutions or authorities} & \makecell[l]{Decentralized control and governance through\\ smart contracts and community participation} \\ \hline

Intermediaries &\makecell[l]{Reliance on intermediaries such as \\banks, brokers, and custodians for\\ transactions and custody of assets} & \makecell[l]{Reliance on intermediaries such as \\banks, brokers, and custodians for\\ transactions and custody of assets} & \makecell[l]{Elimination or reduction of intermediaries \\through P2P network and smart contracts} \\ \hline

\makecell[l]{Transparency \\\& Auditability} & \makecell[l]{Limited transparency and visibility into\\ transactions and processes} & \makecell[l]{Limited transparency and visibility \\into transactions and processes} & \makecell[l]{Transparent and auditable transactions\\ and processes on public blockchains} \\ \hline

\makecell[l]{Access \& \\Financial Inclusion} & \makecell[l]{Access limited by geographical, \\regulatory, and identity barriers} & \makecell[l]{Access limited by regulatory \\requirements and KYC processes} &\makecell[l]{Open access for anyone with an internet \\connection and compatible wallet} \\ \hline

\makecell[l]{Security \\\& Risks} & \makecell[l]{Security relies on trusted intermediaries \\and regulatory frameworks} &\makecell[l]{Security relies on trusted intermediaries \\and regulatory frameworks} & \makecell[l]{Smart contract vulnerabilities, hacking, \\and user errors introduce risks} \\ \hline

\makecell[l]{Scalability} & \makecell[l]{Limited scalability due to centralized \\infrastructure and legacy systems} & \makecell[l]{Limited scalability due to centralized \\infrastructure and legacy systems} & \makecell[l]{Potential for scalability through layer 2 \\solutions and interoperability} \\ \hline

\makecell[l]{Innovation \\\& Flexibility} & \makecell[l]{Slower adoption of new technologies \\and financial products} & \makecell[l]{Innovation driven by centralized \\institutions and their roadmap} & \makecell[l]{Rapid innovation and experimentation in \\developing new protocols and features} \\ \hline
\end{tabular}
}
\caption{Comparison of TradFi, CeFi, and DeFi.}
\label{tab:comparison}
\end{table*}

\emph{\TradFi} refers to the conventional financial system that has been in place for decades. It involves centralized institutions such as banks and regulatory bodies. In \TradFi, intermediaries play a crucial role in facilitating transactions, providing financial services, and managing assets. Transactions are typically conducted through centralized platforms, and the system operates under regulatory frameworks established by governing authorities. \emph{\CeFi} is a model that follows the \TradFi system. It involves financial intermediaries and institutions that act as central authorities in managing and controlling financial transactions and services. \CeFi platforms, such as \CEXs (e.g., \href{https://www.coinbase.com/}{Coinbase}), hold custody of users' crypto-assets and are responsible for executing transactions. These institutions have control over the financial system, including the verification of identities, \AML and \KYC processes, and decision-making. \emph{\DeFi} aims to remove intermediaries, enabling direct \PtP transactions. In \DeFi, transactions are recorded on public blockchains, and users have control over their crypto-assets. \DeFi platforms often utilize governance mechanisms to allow participation in decision-making processes. 

In summary, \TradFi relies on centralized intermediaries to facilitate financial activities, while \DeFi aims to replace these intermediaries with blockchain technology and smart contracts. It is crucial to note that \CeFi is a separate concept, although it shares some similarities with both \TradFi and \DeFi. This paper focuses on the \DeFi ecosystem.

\section{Concerns on \DeFi Market Misconduct} \label{sec: concerns}

\subsection{The bZx Incident}

The \href{https://bzx.network/}{bZx} incident marked a significant event in the \DeFi space. It occurred in February 2020 and was notable for being one of the major exploits leveraging flash loans. The bZx protocol, a decentralized lending and trading platform, was targeted in an attack where an adversary utilized a flash loan to manipulate asset prices and extract significant profits. The attacker borrowed $7{,}500$~\ETH (approximately $2$m~\USD) using a flash loan from dYdX, manipulated the price of \texttt{sUSD} by creating an imbalance in liquidity pools on Uniswap, and exploited this price discrepancy to close a leveraged position on bZx at an artificially high profit.

The attack resulted in losses of approximately $645$k~\USD for the protocol and its users. The incident exposed key vulnerabilities, particularly in the reliance on decentralized price oracles and the misuse of flash loans as a tool for manipulation. The bZx incident served as a wake-up call for the \DeFi community, raising awareness about the potential for market misconduct enabled by innovative \DeFi tools such as flash loans. It underscored the need for stronger risk management practices and regulatory measures to mitigate such risks and protect user assets in the \DeFi ecosystem.

\subsection{Blockchain and DeFi Market Misconduct}
We proceed to answer \textbf{RQ1: How could the misuse or exploitation of blockchain technology exacerbate market misconduct in \DeFi?} By analyzing the potential consequences of improper blockchain use, we aim to highlight how this technology can be exploited to facilitate market misconduct within the \DeFi ecosystem.

\subsubsection{Blockchain Pseudonymity and Mixing Services} \label{sec:mixer}

DeFi platforms often function on smart contract-enabled blockchains, such as Ethereum~(\citealp{wood2014ethereum}), which offer decentralized and pseudonymous participation, allowing transfers without intermediaries. This ecosystem is more likely to attract misconduct activities. In contrast to \TradFi, \DeFi does not require users to link their personally identifiable information to their accounts. Instead, a \DeFi user can adopt a wallet address to receive or send transactions. A wallet address is derived from a cryptographic hash function applied to the user's public key, and the user can control the private key secretly to sign transactions and access the funds. The \emph{unlinkability} between users' personal information and wallet addresses provides malicious actors with greater opportunities to conceal their identities. Furthermore, in contrast to \CeFi where \KYC is widely adopted, establishing such measures on \DeFi protocols poses significant challenges, allowing misconduct activities to evade regulation more easily~(\citealp{moser2019effective, kolachala2021sok}).

Although \DeFi users can achieve pseudonymity on-chain, their transactions are recorded on the public ledger, including the transaction amount, time, and the sender and recipient addresses. The transaction information can be used to cluster and analyze addresses, potentially deanonymizing \DeFi users and revealing their personal identities~(\citealp{ermilov2017automatic, yousaf2019tracing, victor2020address}). To achieve better privacy, on-chain mixers are proposed~(\citealp{wang2023zero}).  Users deposit cryptocurrencies into a mixer pool managed by smart contracts and withdraw to another address. By leveraging Zero-Knowledge Proof (ZKP), users can prove the validity of the deposits and withdrawals without revealing transaction information. Similar to $k$-anonymity, a user of a mixer service can be hidden among of set of other $k$ users, thus breaking the linkability between their deposit and withdrawal addresses.

\begin{figure}[tbh]
\centering
    \includegraphics[width=\columnwidth]{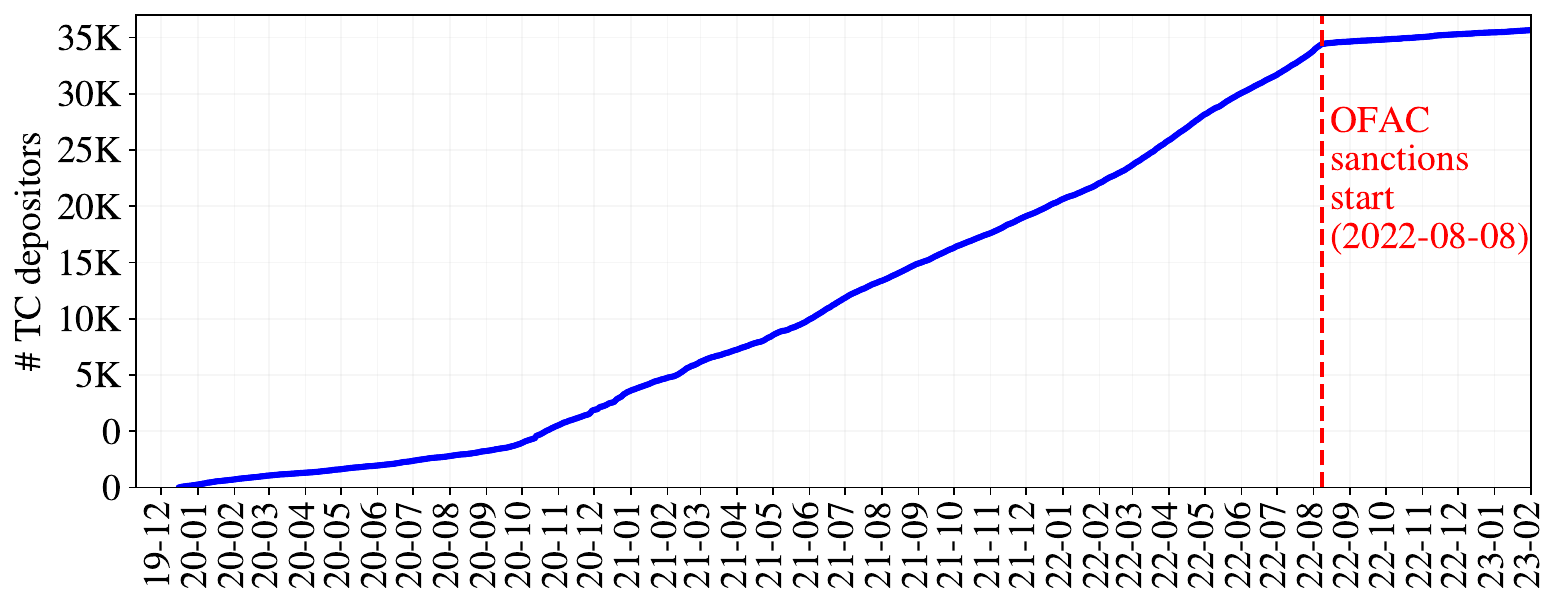}
    \caption{Accumulated number of depositors in TC \ETH pools.}
    \label{fig:mixers_num_depositors_over_time}
\end{figure}

\begin{figure}[tbh]
\centering
    \includegraphics[width=\columnwidth]{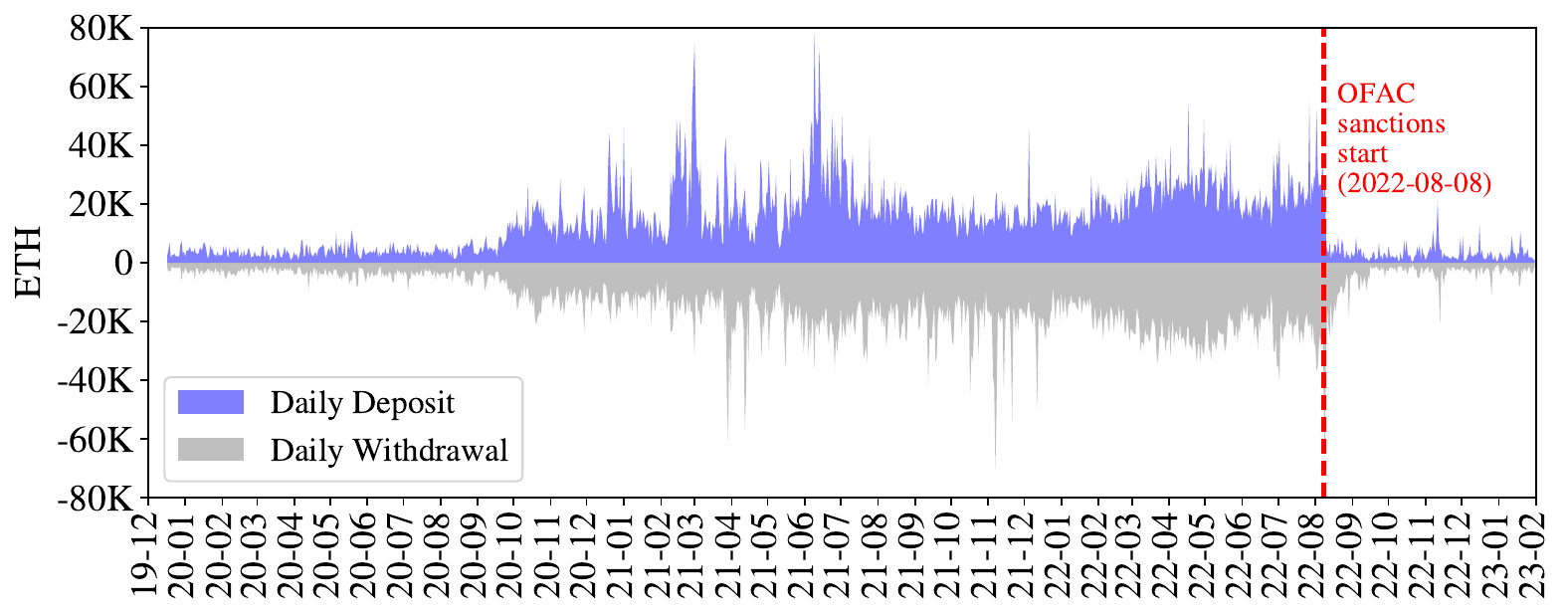}
    \caption{Daily deposit/withdrawal amount in TC \ETH pools.}
    \label{fig:TC_Pools_Daily_activities}
\end{figure}

On-chain mixers, such as \TC, are frequently used by malicious actors to obscure their identities. We identify $40{,}649$ blockchain addresses that have deposited \ETH into the \TC \ETH pools (see Figure~\ref{fig:mixers_num_depositors_over_time} and~\ref{fig:TC_Pools_Daily_activities}), with the total amount exceeding $3.60$m \ETH ($5.40$b \USD). According to \cite{wang2023zero}, at least $412.87$m \USD deposits are from adversaries. \TC's mixing pools obscure the origin of funds, creating challenges in tracking transactions, particularly for law enforcement purposes~(\citealp{wang2023blockchain}). The \OFAC sanctioned \TC in Aug $2022$\footnote{See \href{https://home.treasury.gov/news/press-releases/jy0916}{U.S. Treasury Sanctions Notorious Virtual Currency Mixer Tornado Cash}.}. However, due to the decentralized nature of blockchain, a \DApp like \TC cannot be entirely prohibited by a centralized regulator, allowing sophisticated adversaries to continue interacting with \TC (see~Figure~\ref{fig:TC_Pools_Daily_activities}).

\subsubsection{Blockchain Transparency and Pending Transaction} \label{sec:transparent}

Blockchain operates on a transparent and public ledger, ensuring the visibility and accountability of transactions. This inherent transparency fosters trust and enables the traceability and verification of transactions. Unlike \TradFi users, \DeFi users do not necessarily require inside information to engage in market misconduct. Instead, they can use a well-connected blockchain node to monitor pending and unconfirmed transactions. This provides them with potential opportunities to engage in market misconduct, such as front-running or exploiting time-sensitive information. By acting on pending transactions, malicious \DeFi users can take advantage of market conditions and execute trades ahead of others, potentially harming market fairness and efficiency.

\subsubsection{Blockchain Atomicity and Flash Loans}

\begin{figure}[tbh]
\centering
    \includegraphics[width=\columnwidth]{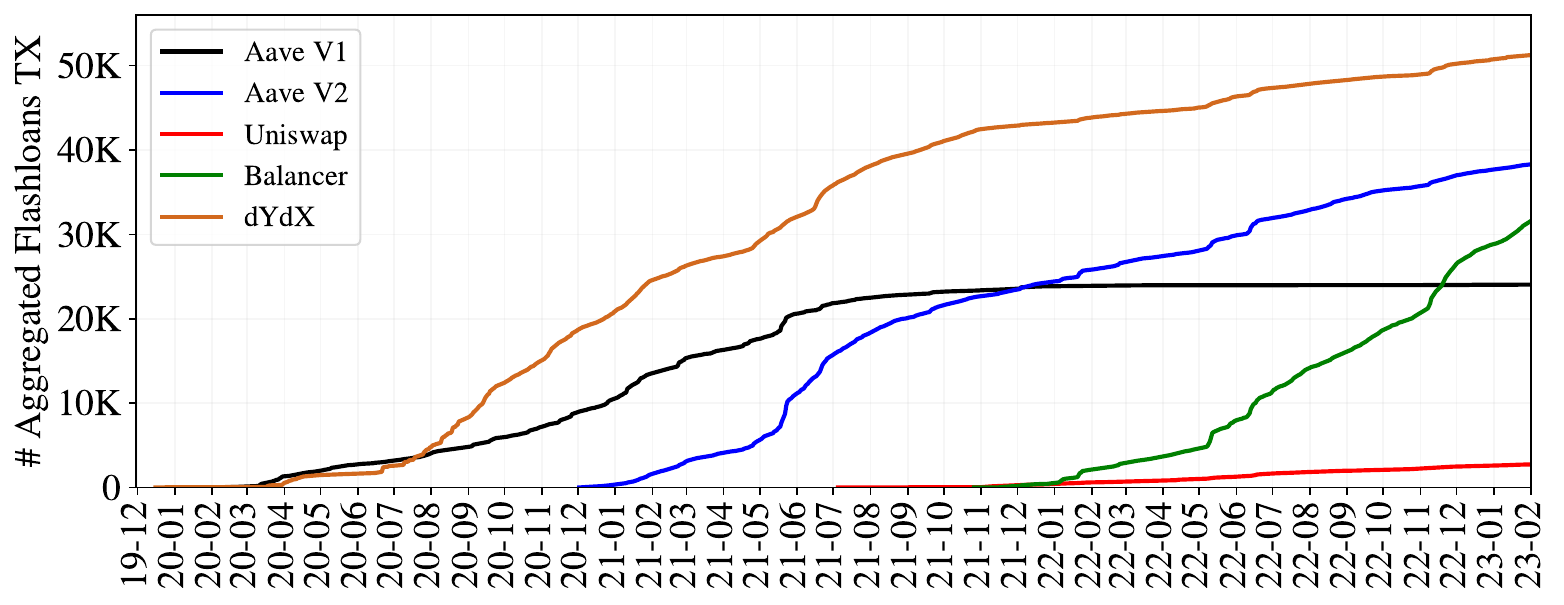}
    \caption{Accumulated Flash loan transactions over time.}
    \label{fig:flashloans_tx_over_time}
\end{figure}

Blockchain atomicity ensures that transactions are executed in an all-or-nothing manner, meaning all actions within a transaction must succeed, or the entire transaction is rolled back, leaving the blockchain state unchanged. This property preserves system integrity by preventing incomplete updates to the ledger. Atomicity also enables innovative \DeFi products like flash loans—a form of zero-collateralized lending where funds are borrowed and repaid within a single atomic transaction. Flash loans provide borrowers with instant access to billions of assets and are commonly integrated with \DeFi platforms such as \href{https://balancer.fi/}{Balancer}, \href{https://app.uniswap.org/swap}{Uniswap}, \href{https://aave.com/}{Aave}, and \href{https://dydx.exchange/}{dYdX}. As noted by~\cite{flashloans}, at least $123.8$k transactions involve flash loans ( Figure~\ref{fig:flashloans_tx_over_time}), totaling $330$b \USD.

Unlike \TradFi, which often demands significant assets to influence price volatility, \DeFi users can leverage flash loans to execute advanced trading strategies and seize opportunities in real-time. 
Flash loans enable users to borrow substantial funds without collateral, provided the loan is repaid within a single atomic transaction. This capability facilitates advanced trading strategies such as arbitrage and liquidity provision, which often yield significant profits. However, the rise of flash loans also introduces new risks. Malicious actors can exploit them to target vulnerable \DeFi protocols and engage in market misconduct. For example, in February 2020, the bZx lending protocol experienced two flash loan attacks, resulting in losses of $950$k \USD.

\subsubsection{DeFi Protocol Composability}

In \TradFi, different services often operate in isolation, making it challenging to combine them to create innovative solutions. However, in \DeFi, protocols are designed to be composable. From the protocol level, composability enables an address specific to a protocol to use addresses from the same or another DeFi protocol to deliver an innovative financial service in a single transaction~(\citealp{10.1145/3532857}). DeFi protocol composability is realized due to the interaction between underlying smart contracts.  For example, when \DeFi users perform $\USDT \rightarrow \KYL$ swap on \href{https://1inch.io/}{1inch} (a \DeFi aggregator platform), 1inch conducts the swap sequentially through two \DEXs using \WETH as an intermediary token, all within an atomic transaction. Upon users interacting with the 1inch smart contract, 1inch triggers the internal transactions on Uniswap for $\USDT \rightarrow \WETH$ and on Sushiswap for $\WETH \rightarrow \KYL$. From the user level, the composability feature enables users to combine various protocols to create trading strategies.

However, the composability feature of DeFi protocols introduces new risks of misconduct within the \DeFi ecosystem. Even a minor vulnerability in the smart contract code can result in protocol failures, potentially impacting other integrated protocols as well. For example, \href{https://coinmarketcap.com/currencies/bzx-protocol/}{bZx} is a DeFi platform offering lending and margin trading services. bZx enables users to contribute capital to pools, borrow against their capital, and engage in leveraged positions using other assets on margin through the power of composability. However, in Feb $2020$, an adversary attacked bZx with flash loans by exploiting the composability between bZx and other platforms, including dYdX, Compound, and Uniswap.

\subsubsection{MEV, Gas Fee Auction and FaaS}

\begin{figure}[tbh]
\centering
    \includegraphics[width=\columnwidth]{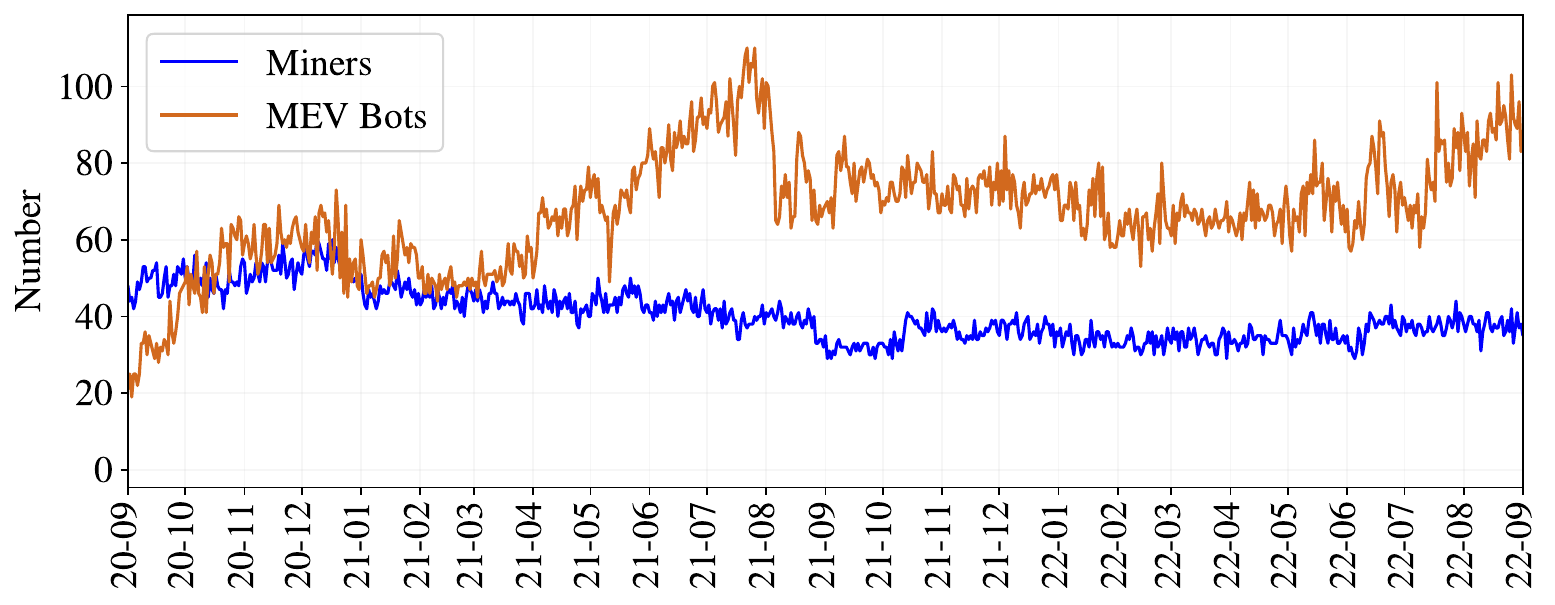}
    \caption{Number of daily MEV bots and miners.
    }
    \label{fig:mev_bots_num_over_time}
\end{figure}

In general, miners and validators have the power to determine the inclusion of the pending transaction in the mempool, as well as their ordering in the block. \cite{daian2020flash} introduce \MEV, which refers to the maximum potential revenue that can be extracted by \DeFi participants through transaction order execution manipulation. Figure~\ref{fig:mev_bots_num_over_time} shows the number of daily \MEV bots and unique miners recorded by Flashbots between Sep~$1$,~$2020$ and Sep~$1$,~$2022$. More than $449$m~\USD \MEV are extracted during this period~(\citealp{Ethereum-MEV-data}).

In \TradFi, bribing brokers is considered illegal. However, the decentralized nature of \DeFi introduces new dynamics where users have the ability to incentivize validators or miners to manipulate the transaction order by participating in gas fee auctions. \MEV searchers, who actively seek out and exploit \MEV opportunities, often participate in auctions for priority transaction inclusion in the block. Participants looking to execute transactions have two options for submission.  They can make their transactions public and submit them to miners or validators via the \PtP network. Alternatively, they have the choice to keep their transactions private and submit them through a centralized \MEV relay known as \FaaS.

First, a trader can auction for \MEV extraction priority by participating in \PGA, in which the trader competitively bids up transaction fees and submits transactions to miners through the public \PtP network in order to obtain priority ordering~(\citealp{daian2019flash}). To participate in an arbitrage opportunity, a trader named $\mathsf{trader}_1$ initiates a transaction $\tx_0$ with a gas price of $gas_0$ and a nonce of $nonce_0$. If $\mathsf{trader}_0$ observes that other participants on the \PtP network are competing for the same arbitrage opportunity by setting a higher gas price of $\hat{gas_0}$, $\mathsf{trader}_0$ has the option to reissue the transaction as $\tx_1$ with the same $nonce_0$ but a higher gas price $gas_1$ (where $gas_1>\hat{gas_0}$) to outbid other participants. Consequently, a bidding war may ensue among multiple traders for the same arbitrage opportunity.

Traders can also extract \MEV by bundling transactions and transmitting them privately to validators via \FaaSs, such as \href{https://boost-relay.flashbots.net/}{Flashbots}. In this network, traders can include multiple signed transactions in their bundles, including transactions issued by other entities. Each bundle contains metadata specifying the execution logic. Instead of broadcasting these bundles over the \PtP network, traders send them directly to miners or validators. Notably, participating in Flashbots auctions is risk-free, as unsuccessful bids incur no transaction fees. Flashbots employs a first-price \SGA mechanism, enabling users to securely submit bids and specify their preferred transaction order. This mechanism maximizes validators' payoffs while providing an efficient platform for \MEV price discovery.

\subsubsection{Smart Contract-Based \MEV Bots}

\MEV bots are on-chain smart contracts designed to identify and exploit \MEV opportunities. Similar to trading bots in \TradFi, \MEV bots are automated programs that perform repetitive tasks more efficiently than human traders. However, as \MEV bots are deployed on-chain, they can easily be composed with existing \DeFi protocols, enabling advanced functionalities such as arbitrage and liquidation.

Another notable aspect of \MEV bots is the transparency of blockchain data, which enables malicious actors to analyze and replicate the strategies of existing \MEV bots, potentially facilitating misconduct. This transparency lowers the barriers to unethical behavior in the \DeFi ecosystem. Historically, once a new format of \DeFi vulnerabilities was reported, it could breed a series of copycat attacks. 
For instance, on the Ethereum blockchain explorer, there are $399$ contract addresses that are labeled as \MEV bots at the time of writing\footnote{\href{https://etherscan.io/accounts/label/mev-bot}{https://etherscan.io/accounts/label/mev-bot}}. Although many \MEV bots conceal their contract code and only display the publicly deployed EVM bytecode on-chain, it is still feasible to decompile the bytecode and analyze the underlying strategies. Furthermore, we have observed instances\footnote{\href{https://etherscan.io/address/0x3156ffa5b02ecc63101581e0610a7f1df5181e1e\#code}{https://etherscan.io/address/0x315...e1e\#code}} where the \MEV bot openly exposes its smart contract code on blockchain explorers, thereby providing opportunities for malicious copy-cats.


\section{Market Misconduct in TradFi} \label{sec: misconduct-tradfi}
In Section~\ref{sec:background}, we examined how blockchain characteristics can facilitate \DeFi market misconduct. This section focuses on the definition and strategies of market misconduct in \TradFi, providing a foundation for comparison with \DeFi market misconduct in Section~\ref{sec: misconduct-defi}.

\subsection{Existing Regulatory Framework}\label{sec:laws}
We begin by analyzing the existing regulatory approaches to market misconduct in major global economies.

\textbf{United States:} In the U.S., market misconduct is regulated through a comprehensive framework. Central to this framework is the \href{https://www.govinfo.gov/content/pkg/COMPS-1885/pdf/COMPS-1885.pdf}{\SEA}, which established the \SEC with broad regulatory authority over securities markets. The \SEC enforces rules such as \href{https://www.law.cornell.edu/cfr/text/17/240.10b-5}{Rule 10b-5}, aimed at preventing fraudulent and manipulative practices. Additionally, the \CFTC oversees commodities and futures markets, implementing regulations under the \href{https://www.govinfo.gov/content/pkg/COMPS-10309/pdf/COMPS-10309.pdf}{\CEA} to combat manipulation and fraud. The \href{https://www.congress.gov/111/plaws/publ203/PLAW-111publ203.pdf}{Dodd-Frank Wall Street Reform and Consumer Protection Act of 2010} further augments this regulatory structure by enhancing corporate governance, transparency, and financial oversight. These laws and regulations create a strong framework to prevent market abuse, ensure transparency, and uphold investor confidence in the financial market.

\textbf{European Union:} In the EU, regulations of market misconduct are harmonized across member states to ensure consistent standards. \href{https://eur-lex.europa.eu/legal-content/EN/TXT/?uri=celex\%3A32014R0596}{\MAR} provides a detailed regime for dealing with insider dealing, market manipulation, and unlawful disclosure of inside information, enhancing the transparency and integrity of financial markets. It is applicable across all EU member states,  including the UK during its EU membership. The European Securities and Markets Authority (ESMA) oversees the enforcement of market abuse regulations across the EU.

\textbf{United Kingdom:} The \href{https://www.legislation.gov.uk/ukpga/2000/8/contents}{\FSMA} established a unified regulatory authority, originally the Financial Services Authority (FSA), which was succeeded in 2013 by the \FCA and the Prudential Regulation Authority (PRA) under the Bank of England. The \FSMA grants the \FCA broad powers to oversee financial markets, enforce market conduct rules, and address market manipulation.
Before Brexit, the UK directly applied the EU's \MAR alongside the \FSMA to address market manipulation and enforce market conduct rules. Following Brexit, the UK retained the core elements of \MAR by incorporating it into domestic law as the UK MAR. 
Together, the \FSMA and \MAR, enforced by the \FCA, provide a robust framework for preventing, detecting, and addressing market manipulation.

\textbf{China:} Market misconduct in China is regulated through the \href{https://www.gov.cn/banshi/2005-07/11/content_13696.htm}{Securities Law}, which addresses insider dealing, market manipulation, and other abusive practices. It prohibits actions that create false or misleading trading conditions, manipulate securities prices, or engage in other fraudulent practices to deceive investors. The law establishes regulations concerning the disclosure of information by listed companies and individuals with insider knowledge. In addition, the Criminal Law and the Anti-Unfair Competition Law also provide provisions to combat market misconduct.

\textbf{Janpan:} Japan's financial market misconduct is primarily regulated by the \href{https://www.fsa.go.jp/en/policy/fiel/index.html}{\FIEA}, which covers areas such as the prohibition of insider trading, the prevention of market manipulation, the enforcement of corporate disclosure, and the registration and conduct standards for financial instruments business operators.  
In Japan, the \FSA ensures the stability of the financial system, while its subordinate body, the Securities and Exchange Surveillance Commission (SESC), focuses on monitoring securities markets.

\subsection{Definition of Market Misconduct}~\label{sec:definition_tradfi}
  
Financial market misconduct comes in various forms. The \MAR prohibits three forms of market misconduct, namely \emph{(i)} market manipulation, \emph{(ii)} insider dealing, and \emph{(iii)} unlawful disclosure of inside information. Market misconduct has been thoroughly researched in a variety of literature, yet it is still challenging to define. The scope of misconduct is extensive, as it includes a diverse array of manipulative strategies~(\citealp{putnicnvs2020overview,Cumming2015financial,cumming2011exchange}).

Despite growing attention to market misconduct, the term remains inadequately defined. The financial and economic literature lacks a precise, concise definition, resulting in an imprecise understanding of the concept. Additionally, legal definitions are often intentionally broad to prevent manipulators from exploiting narrow interpretations to evade the law~(\citealp{putnicnvs2020overview}). In this study, we analyze \TradFi market misconduct using the \MAR framework. 
This approach is chosen because definitions and classifications of market misconduct vary across jurisdictions, and the EU, as one of the largest economic regions, provides a comprehensive and widely referenced regulatory standard through \MAR.

\subsubsection{Market Manipulation}

Market manipulation remains a significant challenge in financial markets~(\citealp{khodabandehlou2022market}). Advancements in technology have made manipulation methods increasingly sophisticated. Consequently, a universally accepted definition of market manipulation has yet to be established~(\citealp{putnicnvs2020overview}). The definition and forms of manipulation continue to be subjects of debate within the academic community and the financial industry.

The work by~\cite{fischel1991should} offers a fundamental analysis of the concept of market manipulation. Their findings suggest that the concept of market manipulation lacks the clarity necessary to serve as a firm basis for criminal charges. They emphasize the absence of an objective definition. Their conclusion implies that manipulation might not be a substantial issue, and they argue against categorizing actual trades as manipulative solely based on the trader's intent. However, \cite{thel1993850} presents a contrasting perspective. They contend that manipulation is not discouraged or unprofitable. Furthermore, they illustrate that manipulators can influence prices through trading and capitalize on short-term price fluctuations. \cite{thel1993850} defines manipulation as trading with the goal of increasing or decreasing asset prices. In the work by \cite{jarrow1992market}, market manipulation is defined as a strategy that generates tangible positive wealth without involving any risk. They investigate whether significant traders possessing market power can manipulate prices in their favor. \cite{cherian1995market} presents a systematic analytical framework for categorizing current research on market manipulation in the equity market. They describe market manipulation as occurring when a manipulator trades with the purpose of intentionally influencing the share price to secure advantages for themselves. In the work by~\cite{kyle2008define}, it is proposed that a trading strategy should be classified as illegal price manipulation if the manipulation is carried out with the intention of implementing a scheme that undermines economic efficiency. This is achieved by diminishing market liquidity for risk transfer and impairing the accuracy of prices as signals for effective resource allocation. \cite{lin2016new} defines market manipulation as a purposeful effort to intervene in market fairness. \cite{eigelshoven2021cryptocurrency} defines market manipulation as trading strategies intentionally devised to undermine market efficiency.

The legal definitions of market manipulation vary across jurisdictions~(\citealp{putnicnvs2020overview}). The \MAR prohibits transactions ``where a person enters into a transaction, places an order to trade or any other behavior which: \emph{(i)} gives, or is likely to give, false or misleading signals as to the supply of, demand for, or price of, a financial instrument, a related spot commodity contract or an auctioned product based on emission allowances; or \emph{(ii)} secures, or is likely to secure, the price of one or several financial instruments, a related spot commodity contract or an auctioned product based on emission allowances at an abnormal or artificial level''\footnote{See Article 12-1(a) of Market Abuse Regulation.}. The \SEA prohibits transactions that ``creates a false or misleading appearance of active trading in any security other than a government security, or a false or misleading appearance with respect to the market for any such security''\footnote{See Section 9-(a)(1) of Security Exchange Act.}.

\subsubsection{Insider Dealing}

It is generally accepted that manipulators can benefit from insider dealing or the release of inside information~(\citealp{allen1992stock}). The \MAR states that inside information shall comprise four types of information. Particularly, for the security market, the inside information is defined as ``information of a precise nature, which has not been made public,..., if it were made public, would be likely to have a significant effect on the prices of those financial instruments or on the price of related derivative financial instruments''\footnote{See Article 7-1(a) of Market Abuse Regulation.}. \MAR defines insider dealing as the situation ``where a person possesses inside information and: \emph{(i)} recommends, on the basis of that information, that another person acquire or dispose of financial instruments to which that information relates, or induces that person to make such acquisition or disposal; or \emph{(ii)} recommends, on the basis of that information, that another person cancel or amend an order concerning a financial instrument to which that information relates, or induces that person to make such cancellation or amendment''\footnote{See Article 8-2(a)(b) of Market Abuse Regulation.}. The SEC \href{https://www.law.cornell.edu/cfr/text/17/240.10b5-1}{Rule 10b5-1} prohibits insider trading cases ``when an individual is trading based on material nonpublic information if that person is aware of the information while engaging in a sale or purchase of securities''.

\subsubsection{Unlawful Disclosure of Inside Information}  

As per the \MAR, the unlawful disclosure of inside information occurs when an individual possesses inside information and discloses that information to another person, which could give recipients an unfair advantage in financial markets\footnote{See Article 10-1 of Market Abuse Regulation.}. However, this prohibition does not apply if the disclosure is made lawfully in the course of regular employment, professional responsibilities, or official duties. \MAR acknowledges that certain situations may necessitate or justify the disclosure of inside information, provided it falls within the scope of legitimate professional obligations.

\begin{table*}[t!]
        \centering
        \resizebox{\linewidth}{!}{ 
        \begin{tabular}{l|lll}
        \toprule
        \makecell[l]{Market\\Misconduct} & \makecell[l]{Type/\\Base} & \makecell[l]{Goal/\\Effect} & Definition\\
        \midrule
        \multicolumn{4}{l}{\greycol{I.Market Manipulation}} \\
        \midrule
        Painting the tape & Trade & \makecell[l]{Price Distortion \\Volume Inflation}& \makecell[l]{Manipulators distort asset prices by buying or selling assets among themselves, which \\often involves wash trades and pre-arranged trades (\citealp{cumming2011exchange}).} \\ 
        
        Wash trade & Trade & \makecell[l]{Price Distortion \\Volume Inflation} & \makecell[l]{An entity trades against itself to create artificial trading volume and increase the price\\ of the target asset~(\citealp{khodabandehlou2022market}).}\\ 

        Pre-arranged & Trade & \makecell[l]{Price Distortion \\Volume Inflation}  &  \makecell[l]{Multiple different but colluding entities buy or sell assets at around the same time with \\the same price and similar volume (\citealp{putnicnvs2020overview}).}\\
        
        Pools & Trade &  \makecell[l]{Price Distortion \\Volume Inflation}  & A group of entities trade assets back and forth among themselves (\citealp{putnicnvs2020overview}).\\

        Churning & Trade & \makecell[l]{Volume Inflation} & \makecell[l]{Excessive buying/selling by a trader (e.g., broker) to generate large commission fees or \\ the appearance of significant volume (\citealp{cumming2011exchange}).}\\

        Ramping & Trade & \makecell[l]{Price Distortion}  & \makecell[l]{The manipulators consecutively execute trades to increase (decrease) prices and \\raise (reduce) demand~(\citealp{cumming2011exchange,khodabandehlou2022market})}.\\
 
        Momentum ignition & Trade & \makecell[l]{Price Distortion}  & \makecell[l]{A series of trades are executed in quick succession to progressively increase (decrease) \\ asset prices, thus inducing short-term trend followers to trade (\citealp{putnicnvs2020overview}).}\\

        Marking the open & Trade & \makecell[l]{Price Distortion} & \makecell[l]{Trading assets at the start of the trading day to alter prices and affect the forecast of\\ the market trend during the day (\citealp{cumming2011exchange,putnicnvs2020overview}).}\\

        Marking the close & Trade & \makecell[l]{Price Distortion} & \makecell[l]{Trading assets shortly before the close of the trading day to alter prices and affect the \\forecast of the market trend for the next day (\citealp{cumming2011exchange,putnicnvs2020overview}).}\\

        Marking the set & Trade & \makecell[l]{Price Distortion} & \makecell[l]{Trading to influence the prices of reference sets (\citealp{putnicnvs2020overview}).}\\

        \makecell[l]{Misleading the\\ month/quarter/year} & Trade & \makecell[l]{Price Distortion} & \makecell[l]{Trading assets on a specific date in an attempt to generate profits or hide losses at \\the end of the month, season, or year (\citealp{cumming2011exchange}).}\\

        Hype-and-dump & Info & \makecell[l]{Price Distortion}   & \makecell[l]{Using the media, the internet, calls/emails, or other channels to disseminate positive, \\false information in an effort to inflate asset prices (\citealp{putnicnvs2020overview}).}\\

        Slur-and-dump & Info & \makecell[l]{Price Distortion}   & \makecell[l]{Using the media, the internet, calls/emails, or other channels to disseminate negative, \\false information in an effort to decrease asset prices (\citealp{putnicnvs2020overview}).}\\

        Capping/pegging & Trade & Price Distortion & Preventing asset prices from rising or falling beyond given thresholds (\citealp{putnicnvs2020overview}).\\

        Action-based  & Action & \makecell[l]{Price Distortion} & \makecell[l]{Taking real actions to affect asset price (\citealp{putnicnvs2020overview}).}\\

        Benchmark rigging  & Submission & \makecell[l]{Price Distortion} & \makecell[l]{Making false or misleading submissions to affect benchmark calculation (\citealp{putnicnvs2020overview}).}\\

        Layering & Order & Volume Inflation & \makecell[l]{Placing one/several orders on one side of a visible limit order book at one or several\\ price steps to mislead the real supply and demand (\citealp{Dalko2020High}).}\\

        Advancing the bid & Order & Price Distortion & \makecell[l]{Raising an asset’s bid price to artificially increase its price~(\citealp{siering2017taxonomy}).}\\

        Reducing the ask & Order & Price Distortion & \makecell[l]{Lowering an asset’s ask price to artificially decrease its price~(\citealp{siering2017taxonomy}).}\\

        Quote stuffing & Order & \makecell[l]{Market Advantage} & \makecell[l]{Submitting and cancelling an enormous number of orders in a short period of time to\\ jam the financial market's infrastructure (\citealp{putnicnvs2020overview,Dalko2020High}).}\\

        Pinging  & Order & \makecell[l]{Spoofing}  & \makecell[l]{Submitting small orders, not intended for execution, in an attempt to discover sizable \\hidden orders and profit from that information (\citealp{putnicnvs2020overview,cumming2011exchange})}.\\

       Giving up priority & Order & \makecell[l]{Spoofing} & \makecell[l]{Submitting a sizable order with a price higher/lower than the best bid/ask price, and\\ cancelling the order as soon as it receives the priority level (\citealp{cumming2011exchange}).}\\

       Switch  & Order & \makecell[l]{Spoofing} & \makecell[l]{Placing a first order on one side and a second order on the other side, and cancelling \\the first order after the second order is completed (\citealp{alexander2022corruption}).}\\

        \midrule
        \multicolumn{4}{l}{\greycol{II.Insider Dealing}} \\
        \midrule
        
        Front-running & Info & Market Advantage & \makecell[l]{Upon receipt of a large client order, a broker trades shortly before a client’s order with\\ the expectation that the client’s order will change the price (\citealp{cumming2011exchange}).} \\
        
        \makecell[l]{Trading ahead of\\research reports} & Info & Market Advantage & \makecell[l]{A broker trades ahead of the release of research reports (\citealp{cumming2011exchange}).}\\
        
        Induced trade & Info & Market Advantage & \makecell[l]{Using inside information to induce another person to acquire or dispose of assets,\\ or cancel/amend an order.}\\
        
        \midrule
        \multicolumn{4}{l}{\greycol{III.Unlawful Disclosure of Inside Information}} \\
        \midrule
        \makecell[l]{Unlawful Disclose} & Info & Market Advantage & \makecell[l]{A person discloses inside information to any other person, except where the disclosure\\is made in the normal course of duties or professional responsibilities.}\\
    
        \bottomrule
        \end{tabular}
        }
        \caption{Taxonomy of TradFi market misconduct under the framework of MAR.}
        \label{tab:tax_tradfi}
    \end{table*}

\subsection{Taxonomy of Market Misconduct}

Table~\ref{tab:tax_tradfi} presents a taxonomy of market misconduct in \TradFi, categorizing them into three types within the context of \MAR: \emph{(i)} market manipulation, \emph{(ii)} insider dealing, and \emph{(iii)} unlawful disclosure of inside information.

\subsubsection{Market Manipulation}

Based on \cite{allen1992stock,putnicnvs2020overview}, we categorize market manipulation into five sub-types based on their mechanisms: \emph{(i)} trade-based, \emph{(ii)} action-based, \emph{(iii)} information-based, \emph{(iv)} order-based, and \emph{(v)} submission-based manipulation. \cite{allen1992stock} explores diverse mechanisms and categorizes market manipulation into three main groups: trade-based, action-based, and information-based manipulation. Building on this classification, \cite{putnicnvs2020overview} proposes two additional categories: order-based and submission-based manipulation.

\emph{Trade-based manipulation} refers to attempts by a trader to influence an asset's value through buying and selling activities, without publicly visible actions to affect the firm's value or spreading deceptive information to impact the asset's price~(\citealp{allen1992stock,chatterjea1993market,menun1993regulation,cherian1995market,nelemans2007redefining,aitken2009trade}). \emph{Action-based manipulation} involves taking concrete actions that directly influence the actual or perceived value of assets~(\citealp{allen1992stock,benabou1992using,chatterjea1993market,cherian1995market,bagnoli1996stock,zhou2003behavior}). \emph{Information-based manipulation} involves the dissemination of false or deceptive information to influence market conditions~(\citealp{allen1992stock,chatterjea1993market,cherian1995market,xin2015preventing}).

\cite{chan2013order} identifies an emerging manipulation mechanism known as \emph{order-based manipulation}, where manipulators create deceptive order displays or artificially complex processing scenarios with no intention of executing the submitted order~(\citealp{Dalko2020High,dalko2018how,chan2013order,kong2014order,putnicnvs2020overview}). 
\cite{Dalko2020High} compares the characteristics of order-based and trade-based manipulation, highlighting a key distinction: order-based manipulation targets the presentation of orders or the processing capabilities of an exchange.
\cite{putnicnvs2020overview} proposes \emph{submission-based manipulation}, which involves submitting deceptive or inaccurate information to influence the calculation of a financial benchmark.

Table~\ref{tab:tax_tradfi} summarizes the market manipulation techniques associated with each of the aforementioned manipulation mechanisms.
For example, \emph{wash trading}, a common form of trade-based manipulation, involves manipulators trading with themselves to artificially inflate trading volume. This creates a misleading perception of market interest in an asset, influencing its price, despite no actual change in ownership~(\citealp{putnicnvs2020overview}). 
Manipulators may conduct wash trades using multiple accounts or brokers to evade detection. \cite{imisiker2018wash} demonstrate that a significant proportion of investors engaged in wash trading on the Istanbul Stock Exchange between $2003$ and $2006$.
The \SEC's \href{https://www.finra.org/rules-guidance/rulebooks/finra-rules/5210}{FINRA Rule 5210} prohibits transactions in securities that result from unintentional order interactions within the same firm, leading to no change in beneficial ownership (e.g., self-trades).

\subsubsection{Insider Dealing}\label{sec:insider}

Insider dealing occurs when an individual in possession of inside information uses it to \emph{(i)} buy or sell financial instruments for their own account or on behalf of clients, or \emph{(ii)} induce another person to engage in such trading.
For example, \emph{front-running} is a common insider dealing technique in which brokers trade financial assets based on advanced knowledge of upcoming transactions expected to impact asset prices.
When a broker receives a significant client order, they may trade shortly before executing the client's order, anticipating that the client's order will impact asset prices~(\citealp{james2018insider}). This practice is explicitly prohibited in \href{https://www.finra.org/rules-guidance/rulebooks/finra-rules/5270}{FINRA Rule 5270}.

\subsubsection{Unlawful Disclosure of Inside Information} 

Unlawful disclosure of inside information involves sharing such information with another person, except when the disclosure occurs in the normal course of duties or professional responsibilities.
\SEC adopts the misappropriation theory of insider trading (\citealp{barczentewicz20blockchain}). 
\href{https://www.law.cornell.edu/cfr/text/17/240.10b5-2}{Rule 10b5-2} prohibits the disclosure of inside information misappropriated in breach of a duty of trust or confidence, whenever: ``\emph{(i)} a person agrees to maintain information in confidence; \emph{(ii)} the person communicating the material nonpublic information ..., such that the recipient of the information knows or reasonably should know that the person communicating the material nonpublic information expects that the recipient will maintain its confidentiality; or \emph{(iii)} Whenever a person receives or obtains nonpublic information from his or her spouse, parent, child, or sibling; ...''.

\section{Market Misconduct in DeFi} \label{sec: misconduct-defi}

In this section, we first provide a definition of market misconduct within \DeFi. Subsequently, we present a taxonomy of \DeFi market misconduct and discuss RQ2 and RQ3 by comparing the market misconduct in \DeFi and \TradFi.

\subsection{Existing Regulatory Framework} \label{sec: existing_defi}

While clear legal frameworks for addressing misconduct in \TradFi markets are established globally (see Section~\ref{sec:laws}), no such frameworks exist for \DeFi markets yet. Currently, no country has established laws specifically regulating \DeFi, let alone legislation targeting misconduct within the \DeFi market. Although the EU has introduced the \href{https://eur-lex.europa.eu/eli/reg/2023/1114/oj}{\MiCA} regulation, which includes provisions for regulating market abuse, \MiCA explicitly states that it does not apply to protocols operating in a fully decentralized manner\footnote{See paragraph (22) of the MiCA regulation.}. This means that while MiCA can offer guidance for addressing market abuse in \DeFi, it is not directly applicable to the regulation of market abuse within \DeFi markets.

\subsection{Definition of \DeFi Market Misconduct}

As outlined in Section~\ref{sec:definition_tradfi}, market misconduct lacks a precise academic or legal definition in \TradFi. This issue is further compounded in the emerging \DeFi sector, where the absence of specific legal frameworks renders the concept even more ambiguous. Hence, our goal is to clearly define \DeFi market misconduct, a crucial step toward achieving conceptual clarity and guiding effective policy development.

\begin{definition}{}\label{def:defi_misconduct}
\emph{\DeFi market misconduct} encompasses actions that exploit vulnerabilities within the \DeFi system, with the intention of undermining market efficiency and/or market fairness, which may pose risks to \DeFi security.
\end{definition}

According to \cite{gilson1984mechanisms,beaver1981market,dimson1998brief}, \emph{market efficiency} refers to the capacity of market prices to accurately and promptly reflect all available information. 
Manipulation can undermine market efficiency by distorting prices, creating artificial demand or supply, or spreading false information that misleads participants. Such actions disrupt resource allocation, hinder the market’s ability to establish fair prices, and impair the decision-making process of \DeFi users.

\emph{Fairness} involves treating all market participants equitably, providing equal access to information, and safeguarding against unfair advantages or disadvantages~(\citealp{kahneman1986fairness,shefrin1993ethics,angel2013fairness}). \DeFi market misconduct undermines fairness by providing certain entities with unfair advantages. This can include inside dealing or other deceptive practices that allow manipulators to gain profits at the cost of others. Such actions distort market competition and compromise fair participation.

\emph{Security} encompasses two key aspects: market security and blockchain security. Market security focuses on safeguarding the assets of \DeFi users within the market environment. This involves measures to protect against risks such as fraudulent activities and hacks. On the other hand, blockchain security involves ensuring the overall safety and integrity of the underlying blockchain system, its protocols, and transaction data. It involves protecting against unauthorized access, disruption, modification, or destruction of the blockchain infrastructure and \DeFi applications. Market misconduct may pose security risks to the \DeFi ecosystem. For example, \MEV searchers often engage in \PGAs to compete for priority transaction ordering by bidding up transaction fees. \cite{daian2019flash} shows that such behavior presents a systemic risk to consensus layer security.

\smallskip Definition~\ref{def:defi_misconduct} specifies that the objective of \DeFi market misconduct is to undermine market efficiency or fairness.
It is important to note that intentional actions targeting the \DeFi infrastructure layer, while potentially impacting \DeFi security, do not necessarily affect market efficiency or fairness and thus are not classified as \DeFi market misconduct.
For example, an adversary may launch an eclipse attack on the NET layer to monopolize connections to a target node within the \PtP network~(\citealp{heilman2015eclipse}). Such an attack poses risks to blockchain security and disrupts consensus but may not directly threaten the \DeFi market. Consequently, attacks affecting blockchain consensus without directly influencing the \DeFi market are excluded from this study's definition of \DeFi market misconduct.


\subsection{Taxonomy of \DeFi Market Misconduct}

\begin{table*}[h!]
\centering
\resizebox{0.99\linewidth}{!}{ 
    \begin{tabular}{l|llllllllll}
    \toprule
    \makecell[l]{Market \\ Misconduct} & \makecell[l]{Type/Base} & \makecell[l]{Goal/Effect} & TradFi & DeFi & \makecell[l]{Efficiency\\Violation}  & \makecell[l]{Fairness\\Violation}  &  \makecell[l]{Security\\Violation}&  \makecell[l]{Layer} &  \makecell[l]{GEV\\Sources} & \makecell[l]{MEV\\Sources} \\
    \midrule
    \multicolumn{11}{l}{\greycol{I.Market Manipulation}} \\
    \midrule

    Marking the open & Trade & \makecell[l]{Price Distortion} & \ding{52} & \ding{56} & N/A & N/A & N/A & N/A & N/A & N/A \\

    Marking the close & Trade & \makecell[l]{Price Distortion} & \ding{52} & \ding{56} & N/A & N/A & N/A & N/A & N/A & N/A \\

    Marking the set & Trade & \makecell[l]{Price Distortion} & \ding{52} & \ding{56} & N/A & N/A & N/A & N/A & N/A & N/A \\
    
    \makecell[l]{Misleading the\\ month/quarter/year} & Trade & \makecell[l]{Price Distortion} & \ding{52} & \ding{56} & N/A & N/A & N/A & N/A & N/A & N/A \\

    Benchmark rigging  & Submission & \makecell[l]{Price Distortion} & \ding{52} & \ding{56} & N/A & N/A & N/A & N/A & N/A & N/A \\
    \hline

    Churning & Trade & \makecell[l]{Volume Inflation} & \ding{52} & \ding{52} & \fc & \hc & \ec & PRO & \ding{56} & \ding{56} \\
    
    Layering & Order & Volume Inflation & \ding{52} & \ding{52} & \fc & \hc & \ec& PRO &\ding{56}  & \ding{56}\\

    Advancing the bid & Order & Price Distortion & \ding{52} & \ding{52} & \fc & \hc & \ec& PRO &\ding{56}  & \ding{56}\\

    Reducing the ask & Order & Price Distortion & \ding{52} & \ding{52} & \fc & \hc & \ec& PRO &\ding{56}  & \ding{56}\\

    Quote stuffing & Order & \makecell[l]{Market Advantage} & \ding{52}  & \ding{52} & \fc & \fc & \fc& PRO/AUX &\ding{56}  & \ding{56}\\

    Pinging  & Order & \makecell[l]{Spoofing}  & \ding{52} & \ding{52} & \fc & \hc & \ec& PRO &\ding{56}  & \ding{56}\\

    Giving up priority & Order & \makecell[l]{Spoofing} & \ding{52} & \ding{52} & \fc & \hc & \ec& PRO &\ding{56}  & \ding{56}\\

    Switch  & Order & \makecell[l]{Spoofing} & \ding{52} & \ding{52} & \fc & \hc & \ec& PRO &\ding{56}  & \ding{56}\\
     
    Painting the tape & Trade & \makecell[l]{Price Distortion \\Volume Inflation}& \ding{52} & \ding{52} & \fc & \fc & \ec& PRO &\ding{56}  & \ding{56}\\

    Wash trade & Trade & \makecell[l]{Price Distortion \\Volume Inflation} & \ding{52} & \ding{52} & \fc & \fc & \ec& PRO &\ding{56}  & \ding{56}\\

    Pre-arranged & Trade & \makecell[l]{Price Distortion \\Volume Inflation}  & \ding{52} & \ding{52} & \fc & \fc & \ec& PRO &\ding{56}  & \ding{56}\\

    Pools & Trade &  \makecell[l]{Price Distortion \\Volume Inflation} & \ding{52} & \ding{52} & \fc & \fc & \ec& PRO &\ding{56}  & \ding{56}\\

    Ramping & Trade & \makecell[l]{Price Distortion}  & \ding{52} & \ding{52} & \fc & \hc & \ec& PRO &\ding{56}  & \ding{56}\\

    Momentum ignition & Trade & \makecell[l]{Price Distortion}  &  \ding{52} & \ding{52} & \fc & \hc & \ec& PRO &\ding{56}  & \ding{56}\\

    Capping/pegging & Trade & \makecell[l]{Price Distortion} & \ding{52} & \ding{52} & \fc & \hc & \ec& PRO &\ding{56}  & \ding{56}\\

    Hype-and-dump & Information & \makecell[l]{Price Distortion}   & \ding{52} & \ding{52} & \fc & \hc & \ec& PRO &\ding{56}  & \ding{56}\\

    Slur-and-dump & Information & \makecell[l]{Price Distortion}   & \ding{52} & \ding{52} & \fc & \hc & \ec& PRO &\ding{56}  & \ding{56}\\

    Action-based  & Action & \makecell[l]{Price Distortion} & \ding{52} & \ding{52} & \fc & \hc & \ec& AUX &\ding{56}  & \ding{56}\\

    \makecell[l]{Front-running} & Transaction & \makecell[l]{Unfair Sequencing}  & \ding{52} & \ding{52} & \hc & \hc & \hc & \makecell[l]{CON/PRO} &\ding{56}  & \ding{52}\\
    \hline

    \makecell[l]{Back-running} & Transaction & \makecell[l]{Unfair Sequencing}  & \ding{56} & \ding{52} & \hc & \hc & \hc & \makecell[l]{CON/PRO} &\ding{56}  & \ding{52}\\

    \makecell[l]{Sandwich} & Transaction & \makecell[l]{Unfair Sequencing}  & \ding{56} & \ding{52} & \hc & \hc & \hc & \makecell[l]{CON/PRO} &\ding{56}  & \ding{52}\\

    \makecell[l]{JIT liquidity attack} & Transaction & \makecell[l]{Unfair sequencing}  & \ding{56} & \ding{52} & \hc & \hc & \hc & \makecell[l]{CON/PRO} &\ding{56}  & \ding{52}\\

    \makecell[l]{Liquidation} & Transaction & \makecell[l]{Unfair sequencing}  & \ding{56} & \ding{52} & \hc & \hc & \hc & \makecell[l]{CON/PRO} &\ding{56}  & \ding{52}\\
 
    \makecell[l]{Oracle manipulation\\(on-chain)} & Transaction & \makecell[l]{Price Distortion}  & \ding{56} & \ding{52} & \fc & \fc & \fc & \makecell[l]{PRO} &\ding{56}  & \ding{56}\\

    \makecell[l]{Oracle manipulation\\(off-chain)} & Data & \makecell[l]{Data manipulation}  & \ding{56} & \ding{52} & \fc & \fc & \fc & \makecell[l]{AUX} &\ding{56}  & \ding{56}\\

    \makecell[l]{Smart contract \\exploitation} & Contract & \makecell[l]{Contract \\manipulation}  & \ding{56} & \ding{52} & \fc & \hc & \fc & \makecell[l]{SC} &\ding{56}  & \ding{56}\\

    \makecell[l]{Governance attack} & Contract & \makecell[l]{Unfair governance}  & \ding{56} & \ding{52} & \fc & \fc & \fc & \makecell[l]{PRO} &\ding{52}  & \ding{56}\\

    \makecell[l]{Rug pull} & Information & \makecell[l]{Scam}  & \ding{56} & \ding{52} & \fc & \fc & \fc & \makecell[l]{AUX} &\ding{52}  & \ding{56}\\ 

    \makecell[l]{Backdoor} & Contract & \makecell[l]{Scam}  & \ding{56} & \ding{52} & \fc & \fc & \fc & \makecell[l]{SC/AUX} &\ding{52}  & \ding{56}\\ 

    \makecell[l]{Honeypot} & Contract & \makecell[l]{Scam}  & \ding{56} & \ding{52} & \fc & \fc & \fc & \makecell[l]{SC/AUX} &\ding{52}  & \ding{56}\\

    \midrule
    \multicolumn{11}{l}{\greycol{II.Insider Dealing}} \\
    \midrule
    
    \makecell[l]{Trading ahead of\\research reports} & Information & Market Advantage & \ding{52} & \ding{56} & N/A & N/A & N/A & N/A & N/A & N/A \\

    Induced trade & Information & Market Advantage & \ding{52} & \ding{56} & N/A & N/A & N/A & N/A & N/A & N/A \\
    \hline
    \makecell[l]{MEV extraction \\against POFs} & \makecell[l]{Information\\Transaciton} & \makecell[l]{Market Advantage\\Unfair sequencing} & \ding{56} & \ding{52} & \hc & \fc & \hc & PRO/CON & \ding{56} & \ding{52} \\
    \midrule
    \multicolumn{11}{l}{\greycol{III.Unlawful Disclosure of Inside Information}} \\
    \midrule
    \makecell[l]{Unlawful Disclose} & Information & Market Advantage & \ding{52} & \ding{52} & \fc & \fc & \ec& AUX &\ding{56}  & \ding{56}\\
        
    \bottomrule
    \end{tabular}
}
\caption{Taxonomy of market misconduct in DeFi and a comparative analysis with TradFi market misconduct. 
\\ '\fc': complete violation; '\hc': partial violation; '\ec': no violation.
}
\label{tab:defi_misconduct}
\end{table*}

Table~\ref{tab:defi_misconduct} presents a comprehensive taxonomy of \DeFi market misconduct and a comparative analysis of misconduct in both \DeFi and \TradFi markets. To the best of our knowledge, we are the first to propose such a taxonomy.

The first part of Table~\ref{tab:defi_misconduct} shows the market misconduct unique to \TradFi. Certain forms of trade-based misconduct, such as manipulating the open/close price, do not exist in \DeFi. This is due to the inherent characteristics of cryptocurrencies, which are traded continuously (i.e., $24$ hours a day and $7$ days a week). The absence of fixed trading hours and closing prices in \DeFi makes it impractical to execute manipulation strategies that rely on specific market timings associated with \TradFi. In the following, we provide a detailed comparison of misconduct in \DeFi and \TradFi.

\subsubsection{Comparable Market Misconduct}

We proceed to address \textbf{RQ2: What forms of market misconduct in \DeFi are comparable to those observed in \TradFi?}  By comparing market misconduct in \DeFi and \TradFi, we aim to analyze forms of misconduct that transcend the boundaries of these two domains.

\noindent\textbf{Order-Based Manipulation}. We first discuss order-based market misconduct in the context of \DeFi. While the majority of \DeFi \DEXs operate using Automated Market Maker (AMM) mechanisms that rely on predetermined mathematical formulas and algorithms, there exist order-based \DEXs such as the \href{{https://0x.org/}}{0x protocol} and \href{{https://idex.io/}}{IDEX}. These platforms introduce the potential for order-based misconduct in \DeFi. 
In detecting order-based market misconduct on \DeFi platforms, several challenges arise. Beyond the inherent characteristics of blockchain—such as pseudonymity, decentralization, and the transparency of on-chain data—the most significant difficulty lies in the off-chain nature of order matching in order-book-based \DEXs. Unlike AMM-based \DEXs, where all transactions are fully executed on-chain, order-book-based \DEXs handle the matching process off-chain, leaving only partial data recorded on the blockchain. This off-chain processing obscures critical details, such as the full order book, canceled orders, and unmatched bids, making it challenging to detect manipulative activities such as spoofing or layering.

\noindent\textbf{Trade-Based Manipulation}. We further examine trade-based manipulation and provide examples for discussion.

\textbf{Wash Trading}: Wash trading is a widely recognized form of misconduct in \TradFi markets, which is prohibited and considered illegal in many jurisdictions. 
This practice involves collusion among multiple actors who conduct artificial trades among themselves, creating the appearance of buying and selling activity while maintaining their positions without assuming any real market risk. A single user can achieve a similar effect by operating multiple accounts.

In \TradFi, information about individual accounts involved in wash trading is typically not publicly accessible. In contrast, \DeFi lowers the barriers for malicious actors to engage in wash trading. Blockchain addresses, equivalent to user accounts in \TradFi, can be created at no cost and without requiring personal identity information. This contributes to the pseudonymous nature of \DeFi transactions, enabling users to operate with a degree of privacy. These reduced entry barriers make wash trading easier to execute, increasing its prevalence in the \DeFi space.
For instance, the findings in \cite{victor2021detecting} reveal that wash trading activities amounting to a sum of $159$m \USD has been identified on two Ethereum limit order book-based \DEXs, namely \href{https://idex.io/}{IDEX} and \href{https://twitter.com/etherdelta?lang=en}{Etherdelta}, between Sep~$2$,~$2017$ and Apr~$5$,~$2020$. \cite{victor2021detecting} also indicates that more than $30\%$ of the investigated tokens have been subjected to wash trading.

\textbf{Pump-and-Dump}:The pump-and-dump scheme, a well-known form of market misconduct in \TradFi, has also emerged in \DeFi. This scheme involves manipulative actors artificially inflating the price of a specific cryptocurrency to trigger a surge in buying activity. Once the price peaks, these actors sell off their holdings, leaving other investors with substantial losses. The decentralized nature of \DeFi exacerbates concerns about the proliferation of pump-and-dump schemes, as malicious actors can exploit the absence of regulatory oversight.
According to \cite{xu2019anatomy}, between Jun $17$,~$2018$, and Feb $26$,~$2019$, approximately $100$ organized Telegram pump-and-dump channels coordinated an average of two pumps per day. These activities resulted in an accumulated trading volume of $6$m \USD per month.

\textbf{Front-running}: Front-running is also observable in the \DeFi space.  In \TradFi, it refers to the practice of gaining an advantage by accessing market information about upcoming transactions before others (see Section~\ref{sec:insider}). For example, a malicious broker may front-run by purchasing a stock for personal gain after receiving a client’s purchase instructions but before executing the transaction.

While \TradFi front-runners rely on inside information to engage in front-running, \DeFi front-runners, on the other hand, may not necessarily require such privileged information.  Front-running in \DeFi happens when the adversary detects a target transaction ($\tx_T$) that is pending on the \PtP network and subsequently capitalizes on this information by submitting its own transaction ($\tx_A$) ahead of $\tx_T$. Since pending transactions are publicly accessible, the adversary can monitor a victim's incoming transaction, construct a new transaction, and pay a higher fee to ensure its placement ahead of the victim's transaction. 
For example, consider a smart contract that offers a reward to a user who manages to guess the preimage of a hash correctly. In this scenario, an adversary can wait until a user uncovers and submits the solution to the network. Upon observing this solution, the adversary duplicates it and engages in front-running. Consequently, the adversary's transaction gains priority, allowing them to claim the reward. Conversely, the user's transaction might be included later, leading to a potential failure~(\citealp{torres2021frontrunner}). This scenario is known as the ``generalized front-running attack''~(\citealp{darkforest2020,baum2021sok}), in which an adversary detects a profitable pending transaction ($\tx_T$) and replicates it using their own account, thereby depriving the original party of potential profits.

\subsubsection{Emerging Market Misconduct} \label{sec:defi_unique}

Next, we focus on \textbf{RQ3: What novel forms of misconduct have emerged in the \DeFi market, facilitated by its unique characteristics and decentralized nature?} Our objective is to analyze the new types of market misconduct arising within this ecosystem.

\textbf{Back-running:} In \DeFi, back-running happens when an adversary observes a pending victim transaction with a significant impact on the market, such as a large purchase or sale order, and intentionally places its own transaction after the victim transaction to capitalize on the anticipated price movement. The technique is often observed in situations where there is an advantage in being the first to execute a transaction following a blockchain state change. Note that the practice of back-running is commonly linked with arbitrage opportunities. For instance, consider a scenario where a pending transaction, $\tx_T$, has the potential to cause a price discrepancy for an asset pair between two \DEXs. Upon observing $\tx_T$ in the mempool, the adversary submits transaction $\tx_A$ with a higher gas fee, anticipating that $\tx_A$ will be executed immediately after $\tx_T$. 

It is worth noting that \TradFi users may also respond to the price discrepancy by engaging in arbitrage, aiming to profit from being the quickest to react. In \TradFi, arbitrage is both legal and encouraged as it facilitates the integration of new information into market prices. 
However, in the \DeFi market, users who engage in arbitrage through methods such as back-running may face legal ambiguity, as back-running is a strategy tied to MEV extraction. The uncertain legality of \MEV extraction (see Section~\ref{sec:mev}) raises concerns about the legitimacy of such arbitrage activities, potentially exposing users to financial market misconduct.

\begin{figure}[tbh]
\centering
    \includegraphics[width=0.95\columnwidth]{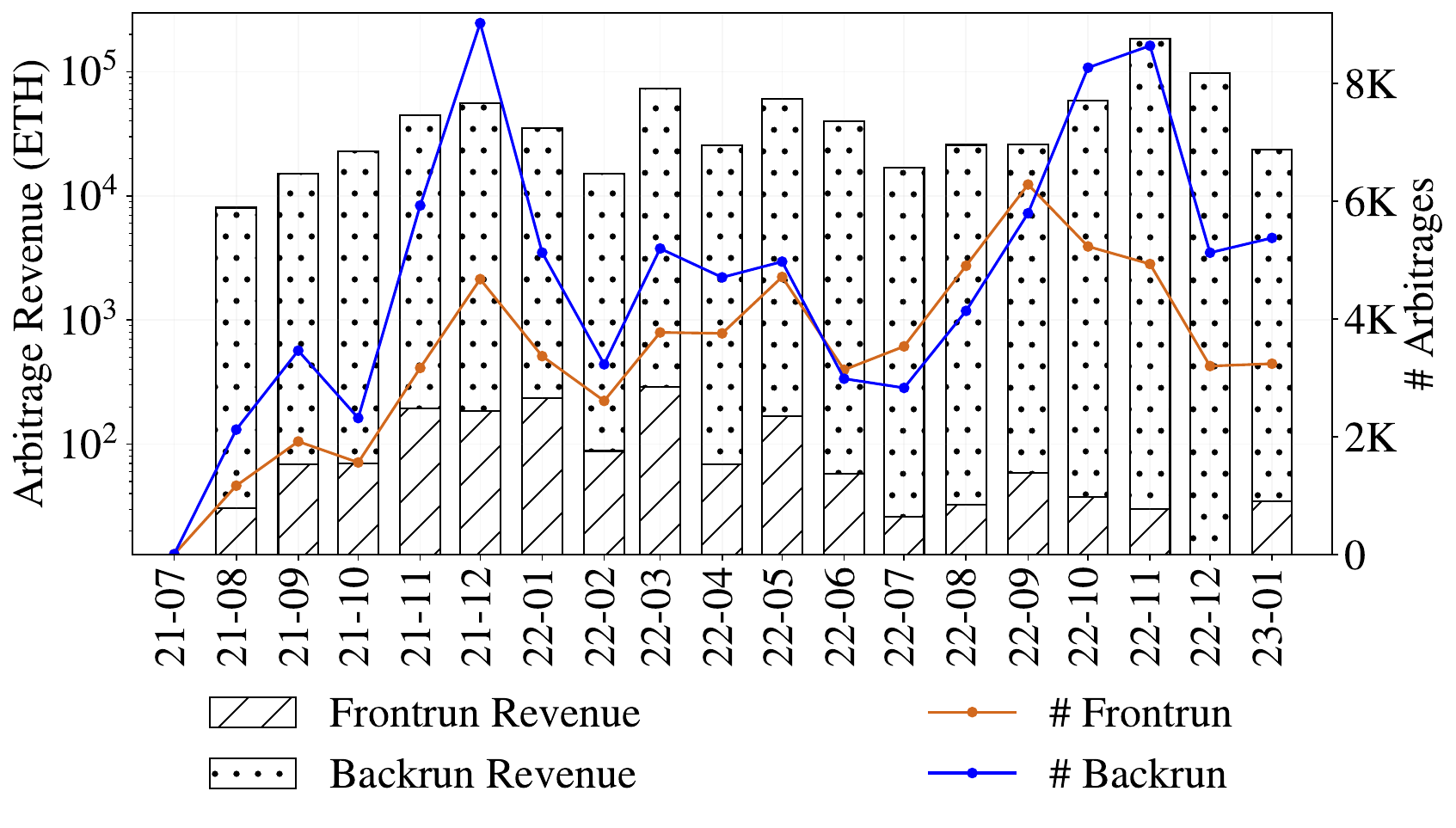}
    \caption{Distribution of detected arbitrages.}
    \label{fig:arbitrage_eth_amount_over_time}
\end{figure}

We empirically analyze the Ethereum arbitrage trades on \href{https://app.uniswap.org/swap}{Uniswap}, \href{https://www.sushi.com/}{Sushiswap}, \href{https://docs.curve.fi/protocol-overview/}{Curve}, \href{https://1inch.io/}{1inch}, and \href{https://bancor.network/}{Bancor}. Figure~\ref{fig:arbitrage_eth_amount_over_time} shows the distribution of captured arbitrages on Ethereum during the period from \dateStart to \dateEnd. Our analysis reveals that a total of $\empirical{1{,}365}$ arbitrageurs with $\empirical{154{,}491}$ transactions, resulting in a cumulative revenue of $\empirical{827{,}376}$ \ETH ($1.16$b \USD). The average gas fee for the captured arbitrage transactions is $\empirical{0.034}$ \ETH ($\empirical{48}$ \USD). Among the $\empirical{154{,}491}$ arbitrage transactions, $\empirical{89{,}281}$ ($57.79\%$) are classified as back-running arbitrages. Interestingly, we find that these back-running arbitrages generate revenue of $\empirical{825{,}687}$ \ETH, representing $99.80\%$ of the total detected revenue.

\textbf{Sandwich attack:} A sandwich attack happens when an adversary strategically places two transactions around a target transaction in the mempool. The adversary listens to the \PtP network to identify a pending transaction ($\tx_T$). Then the adversary quickly submits a transaction ($\tx_{A_1}$) to either purchase or sell the same asset as $\tx_T$, but at a slightly better price. This transaction is designed to front-run the target transaction. Once $\tx_{A_1}$ is included in a block, the adversary ensures that $\tx_T$ follows closely.  Immediately after $\tx_T$, the adversary executes a second transaction ($\tx_{A_2}$) in the opposite direction of $\tx_{A_1}$. This is done to capitalize on the price movement caused by the target transaction. Just like front-runners, the sandwich adversaries may engage in \PGAs or \SGAs to compete against others for the \MEV opportunity. Interestingly, \cite{wang2022impact} uncovered that many users were unaware of sandwich attacks due to their limited technical expertise.

\begin{figure}[tbh]
\centering
    \includegraphics[width=0.95\columnwidth]{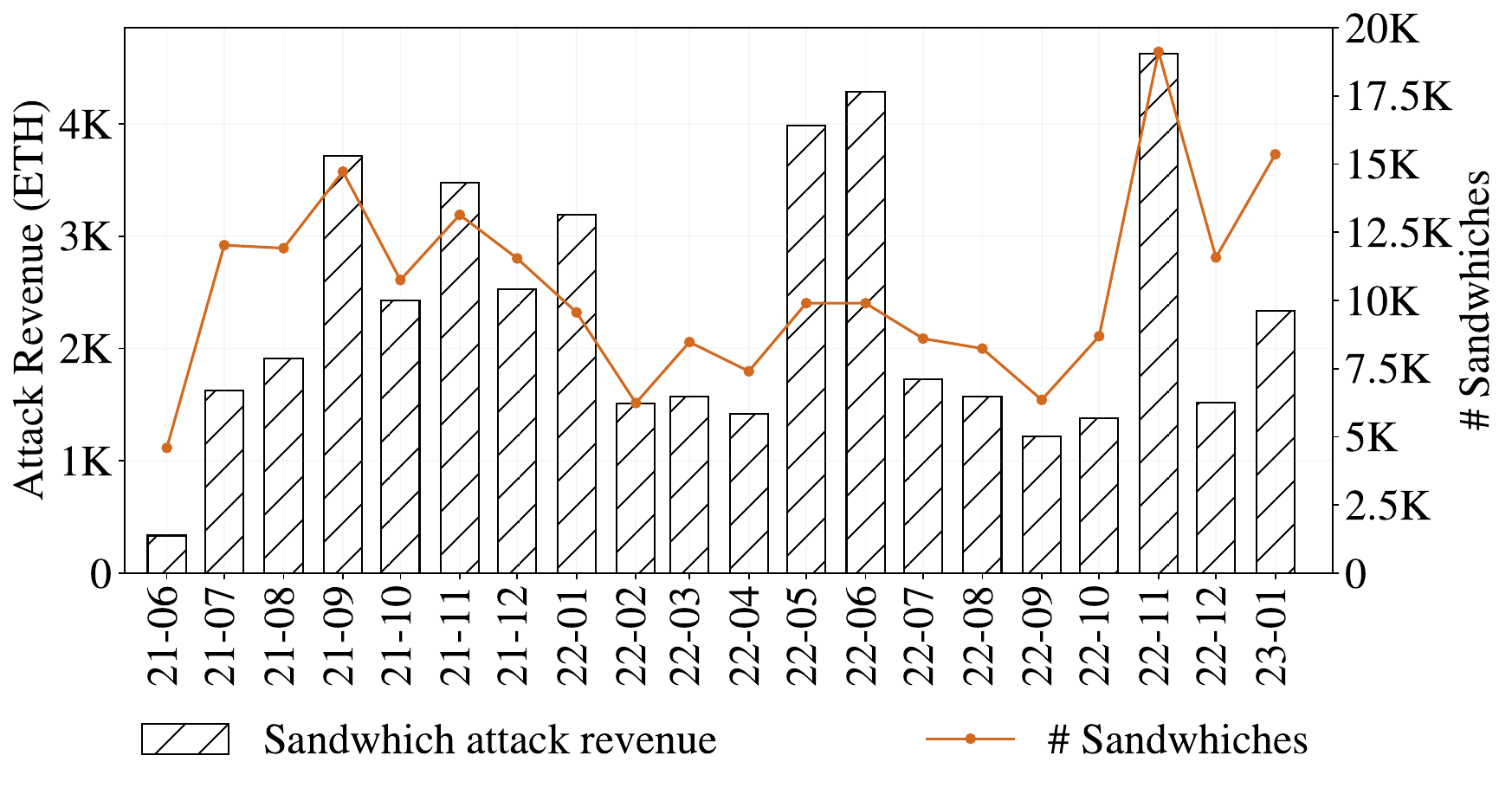}
    \caption{Distribution of detected sandwich attacks.}
    \label{fig:sandwhich_eth_amount_over_time}
\end{figure}

We conduct an empirical study of sandwich attacks on Uniswap V3 from \dateStart to \dateEnd. We identify a total of \numSandwich sandwich attacks, occurring at a monthly frequency of $10{,}407$ attacks (cf.\ Figure~\ref{fig:sandwhich_eth_amount_over_time}). These attacks collectively generate a profit of \totProfitSandwich \ETH ($17.14$m \USD), with an average profit per attack of \avgProfitSandwich \ETH. We observe that sandwich attacks have relatively low entry barriers, with an average initial capital requirement of only \capitalInvSdw \ETH. This amount averages \swapRatioSdw$\times$ higher than the swap volume, implying a relatively small investment compared to the trading activity. Besides, we find that the average \ROI is \avgROISandwich, suggesting profitable attack outcomes. Moreover, we discovered that attackers are open to allocating \bribeRatioSandwich of their attack revenues towards bribing the miners/validators. This indicates a strategic decision to invest a portion of their profits to ensure favorable transaction ordering.

\textbf{\JIT liquidity attack:} This type of attack has become feasible due to the concentrated liquidity feature introduced by Uniswap V3, which empowers \LPs to focus their liquidity within customized price ranges. However, this design also leads to \JIT liquidity attacks, where a malicious actor mints and burns a new position immediately before and after a swap transaction, gaining an unfair advantage in the process~(\citealp{xiong2023demystifying}). An adversary first listens to the public \PtP network to find a sizable target swap $\tx_T$. Upon observation, the adversary calculates a specific price range within which $\tx_T$ is expected to fall, and just before $\tx_T$ occurs, it initiates $\tx_{A_1}$ to create a new liquidity position. This involves adding an extensive amount of liquidity to the pool. The adversary then issues $\tx_{A_2}$ immediately after $\tx_T$ to burn its liquidity position. By launching a \JIT liquidity attack, the adversary dilutes the existing \LP's liquidity share in the pool, thus earning unfair swap fees. Furthermore, the adversary can also gain profits from the change in their portfolio value. 

While sandwich attacks and \JIT liquidity attacks may appear similar on the surface, they involve different attack mechanisms. We investigate \JIT liquidity attacks on Uniswap V3 following heuristics proposed by \cite{xiong2023demystifying}. Our analysis revealed a total of \numJITs \JIT liquidity attacks from \dateStart to \dateEnd. We find that to initiate a \JIT liquidity attack against a specific transaction $\tx_T$, the attacker is required to provide a substantial amount of liquidity. On average, this liquidity amount is approximately \swapRatio$\times$ greater than the swap volume of $tx_T$. In addition, our findings reveal that \JIT liquidity attacks yield low returns, with an average \ROI of only \avgROI and an average profit of merely \avgProfitETH \ETH. Moreover, our analysis reveals that a single bot, \firstBot, is the dominant player in the \JIT game. \firstBot alone has accounted for $76\%$ of all \JIT attacks, issuing a total of $27{,}983$ attacks and generating a profit of $6{,}900$ \ETH (\firstBotProfitPCT). This emphasizes the centralization of power and resources within the control of a limited number of large-scale players.

\textbf{Liquidation:} In \TradFi, liquidations typically occur in margin trading or futures contracts. In these cases, if a borrower's account falls below a specific margin requirement or fails to meet contract obligations, the lender or exchange has the right to liquidate the borrower's positions to cover the outstanding debt. Liquidations in \TradFi serve as a risk management tool to protect lenders from potential losses. 

While \TradFi liquidations are often facilitated by centralized institutions with established procedures and oversight, \DeFi lending protocols typically rely on decentralized governance and automated smart contracts to execute liquidations. \DeFi liquidation occurs when collateralized assets fall below a required threshold, triggering liquidators to repay a portion of the borrower's debt in exchange for acquiring the collateral at a discount. 
All \DeFi participants can initiate the liquidation process, which involves repaying the debt asset and acquiring the underlying collateral at a discounted price.

\begin{figure}[tbh]
\centering
    \includegraphics[width=\columnwidth]{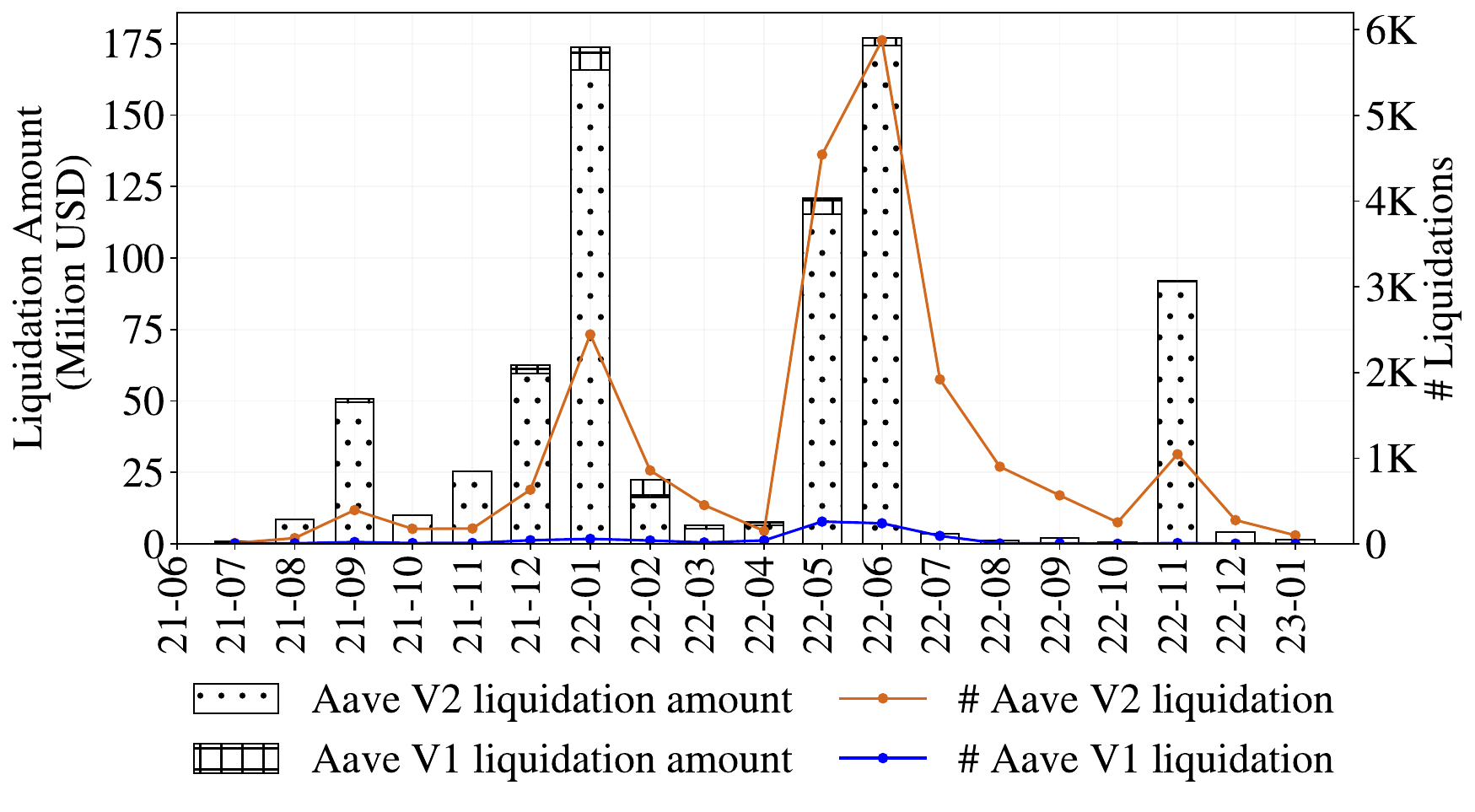}
    \caption{Monthly liquidations occurred on Aave.}
    \label{fig:liquidation_usd_amount_over_time}
\end{figure}

Although \DeFi liquidation plays a crucial role in ensuring the stability of lending platforms, it is vulnerable to exploitation by \MEV searchers. For example, liquidators can listen to the \PtP network and track pending transactions. When they identify transaction $\tx_T$, which may trigger a liquidation opportunity, they can back-run the price oracle update transaction to capture liquidation \MEV. We investigate the liquidations that occurred on Aave V1 and V2. Figure~\ref{fig:liquidation_usd_amount_over_time} shows the monthly liquidations that occurred on Aave from Feb~$1$, 2021 to Jul~$1$, 2023. We find that $\empirical{217}$ liquidators performed $\empirical{21{,}706}$ liquidations during this timeframe, with a total liquidation amount of $\empirical{771}$m \USD. On average, each liquidator captured $100$ liquidation opportunities, corresponding to $\empirical{3.55}$m \USD. The average gas fee is $\empirical{0.09}$~\ETH ($\empirical{132}$~\USD).

\textbf{Price oracle manipulation:} Oracle manipulation is often used as a part of a more comprehensive protocol attack. By manipulating price oracles, the adversary can cause a protocol’s smart contracts to execute based on erroneous inputs, thereby exploiting the situation to their advantage. In blockchain systems, price oracles serve as bridges, linking on-chain smart contracts with off-chain real-world data. For instance, \DeFi protocols often require external information, such as asset price quotes, to function effectively. Money markets, for example, rely on accurate and timely asset prices to issue loans and liquidate collateral appropriately. However, obtaining reliable financial market data poses challenges for \DeFi applications, as blockchains are inherently isolated from external systems, and most financial data is generated off-chain. To address this, blockchain price oracles transmit external data to on-chain smart contracts.

While price oracles are essential for enabling smart contracts to interact with the real world, they are also vulnerable to manipulation and exploitation. Oracle manipulation occurs when an oracle inaccurately reports data about an event or the external world, either due to deliberate actions, negligence, or the compromise of its data source. By exploiting a price oracle, adversaries can deceive smart contracts into executing unfavorable transactions, triggering unintended outcomes, or granting unfair advantages. For instance, adversaries can manipulate oracles to distort asset prices, influence liquidity pools, or exploit arbitrage opportunities for personal gain.
According to \cite{zhou2022sok}, there are at least $30$ oracle manipulation attacks against \DeFi protocols from Apr~$30$,~$2018$ to Apr~$30$,~$2022$, resulting in a total loss of $1.6$b \USD. Oracle manipulation is often part of a broader attack on a \DeFi protocol.  Section \ref{sec:case study} offers a detailed case study regarding price oracle manipulation.


\textbf{Smart contract exploitation:} Given that smart contracts deployed on the blockchain are publicly accessible, any vulnerabilities within these contracts can be exploited by malicious actors. This exploitation may result in unauthorized access, manipulation of the contract's operations, or the misappropriation of funds from \DeFi users or protocols. According to \cite{chaliasos2023smart}, \DeFi attacks targeting smart contracts are increasing, resulting in a total loss of at least $6.45$b \USD. Common smart contract vulnerabilities include reentrancy, integer overflow/underflow, unchecked calls, unhandled error, unchecked math, non-restricted parameters, erroneous visibility, etc. These vulnerabilities, when exploited by malicious actors, can allow them to execute manipulative strategies. For example, vulnerabilities such as reentrancy attacks can be leveraged to manipulate token prices, drain liquidity from pools, or exploit flash loan mechanisms. Reentrancy occurs when contract A calls contract B, and B calls back into A before A finishes its execution. The adversary can drain assets from the victim contract because the reentrant calls will temporarily make the state variables of the caller contract unchanged. A remarkable example is the Lendf.me hack\footnote{See \href{https://peckshield.medium.com/uniswap-lendf-me-hacks-root-cause-and-loss-analysis-50f3263dcc09}{Uniswap/Lendf.Me Hacks: Root Cause and Loss Analysis}.} on Apr $19$,~$2020$, where $25$m \USD worth of funds were stolen from liquidity pools.

\textbf{Governance attack:} A governance attack occurs when malicious actors seek to manipulate the governance mechanisms of \DeFi protocol. These protocols often rely on decentralized governance structures, allowing token holders and stakeholders to vote on decisions such as protocol upgrades, parameter adjustments, and fund allocations. The attacker captures governance extractable value (GEV) by exploiting governance mechanisms to generate economic gains. 
Governance attacks often follow a clear pattern: the adversary first manipulates their voting power, then uses it to approve a malicious proposal to steal funds from the protocol. There are several methods for manipulating voting power. One common approach is leveraging flash loans to temporarily acquire a large number of tokens, granting significant influence over governance proposals. This allows the adversary to implement malicious protocol changes or divert funds. Additionally, the adversary can create multiple fake identities or accounts to gain disproportionate voting power. By controlling a large number of tokens across these fake accounts, the attacker can manipulate the voting process and influence the decision-making process to align with their objectives.
For example, the \href{https://bean.money/}{Beanstalk} protocol suffered a governance attack on Apr~$16$,~2022\footnote{See \href{https://rekt.news/beanstalk-rekt/}{Beanstalk - REKT}.}. The attacker successfully executed a malicious governance proposal and voted to redirect all assets to their own account.

\textbf{Rug pull:} A \DeFi project rug pull refers to an exit scam where the project developer or participants execute a fraudulent exit strategy, causing significant financial losses to investors or users. A rug-pull occurs when a project, often appearing legitimate, exploits trust, marketing tactics, or false promises to attract users to attract investments and gain user deposits before abruptly ceasing operations, taking the capital and disappearing. According to Beosin, in May $2023$, the total amount involved in rug pulls reached $45.02$m \USD\footnote{\url{https://twitter.com/Beosin_com/status/1664171529419243527}}. The most significant case was the alleged $32$m \USD taken by the \DeFi project Fintoch on May~$24$, $2023$. The impact of a rug pull extends beyond immediate financial losses and can affect investor confidence, project reputations, regulatory actions, industry practices, and user awareness.

\textbf{Backdoor/Honeypot:} A smart contract backdoor refers to a hidden or intentionally inserted vulnerability within a smart contract that allows unauthorized access, control, or manipulation by an attacker. These high-privileged functions serve as an entry point to exploit the contract's functionality or steal assets from the contract.  Backdoors may be deliberately introduced by the contract developer or unintentionally created through programming errors or security flaws. When exploited, the backdoor allows the attacker to bypass security measures and gain unauthorized control, potentially resulting in financial losses for the protocol.

A smart contract honeypot is created to attract users by promising high returns opportunities, typically requiring them to transfer a certain amount of crypto assets into the contract upfront~(\citealp{torres2019art}). However, hidden within its code are vulnerabilities or malicious logic that enable the contract creator to exploit users and steal their funds. For example, the contract may manipulate token balances, execute arbitrary transactions, or trap users' funds within the contract. A honeypot may also exploit vulnerabilities in other \DeFi protocols to further deceive and harm users.

Backdoors and honeypots are common techniques in conducting rug pulls~(\citealp{ma2023pied}). Exploitation typically occurs in two ways. First, a malicious contract owner or developer may use backdoors to manipulate trading rules and extract profits. The adversary can leverage hidden vulnerabilities or malicious logic in honeypot contracts to perform unauthorized actions, such as price manipulation for personal gain.
Second, if an adversary gains access to the private key of the contract owner, they can exploit the backdoors to harm the \DeFi protocol. With the private key, the adversary can impersonate the contract owner and execute malicious actions through the backdoors. This may include stealing funds, manipulating data, or disrupting the protocol’s normal operations.
For example, PeckShield\footnote{\url{https://twitter.com/PeckShieldAlert/status/1659404608685604864}} reported that malicious actors stole $3$m \USD worth of \ETH from Swaprum's liquidity pools. The deployer exploited the \texttt{add()} backdoor function to seize staked liquidity provider tokens and withdraw liquidity for personal gain. This attack was enabled by a malicious upgrade of the liquidity reward contract, which intentionally incorporated backdoor functions.

\textbf{MEV extraction against POFs.} As discussed in Section~\ref{sec:transparent}, \MEV extraction in transactions transmitted through public \PtP networks does not rely on private information, as any \MEV searcher operating a node can monitor transactions in the mempool.  However, users (i.e., transaction senders) seeking greater privacy may route their transactions directly to validators using \FaaS services, a practice known as the \POF. This private communication grants \FaaS validators privileged access to inside information about user transactions. In such cases, \FaaS validators can extract \MEV by analyzing the trading intent behind pending transactions. This advantage stems from their ability to either access trading intent earlier than others or leverage superior information for analysis.

As highlighted by~\cite{barczentewicz20blockchain}, the legal risks associated with extracting \MEV from \POFs are significantly higher. A primary concern is the potential for insider dealing, as validators receiving \POFs have access to privileged information that can be exploited for personal gain. As a result, users opting for the \POF may expose themselves to greater vulnerability to \MEV extraction. 
In cases where \MEV extraction involves \POFs, liability may arise if validators engage in misconduct. The use of privileged information and the potential for unfair advantage raise concerns about market abuse and violations of anti-manipulation rules.
Hence, regulators may need to consider these implications and explore ways to mitigate such risks.

\subsection{Case Study} \label{sec:case study}
In the following, we provide a concrete case of \DeFi misconduct:  the price oracle manipulation against an Ethereum \DeFi platform, \href{https://www.harvest.finance/}{Harvest.Finance}, happened on Oct $26$, $2020$. Harvest.Finance is a \DeFi aggregator platform that helps users discover yield-generating opportunities. The assets inside the Harvest.Finance USDC and USDT vaults are deposited into shared pools of underlying \DeFi protocols, i.e., the Y pool on Curve.fi. The assets held in these Harvest.Finance pools are exposed to various market dynamics, including impermanent loss, arbitrage opportunities, and slippage. Consequently, the prices of these assets can be influenced or manipulated through market misconduct, which involves high-volume trades.

\begin{figure*}[tbh]
\centering
\includegraphics[width=0.9\textwidth]{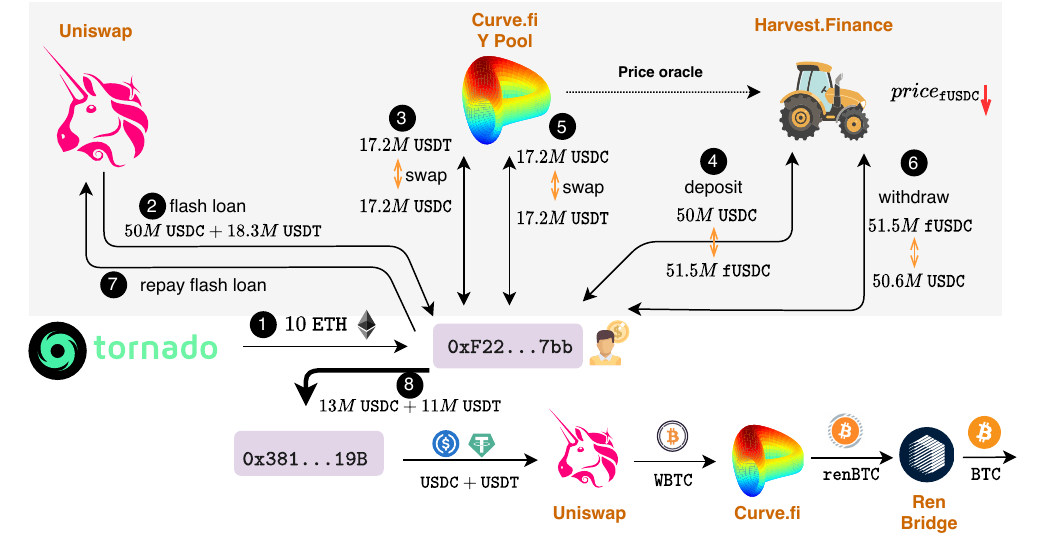}
\caption{\DeFi misconduct case study: the price oracle manipulation against Harvest.Finance, on Oct $26$,~$2020$. The attacker first withdrew $10$~\ETH from the on-chain mixer, Tornado.cash, as the initial fund (Step~\protect\bcircled{1}). The attacker then executed $17$ attack transactions targeting the Harvest.Finance USDC vault, which leveraged flash loans to manipulate the price oracle pended with Curve.fi Y pool, within $4$ minutes (Steps~\protect\bcircled{2} -~\protect\bcircled{7}), and $13$ transactions targeting the USDT vault within another $3$ minutes. The attack yielded $13$m \USDC and $11$m \USDT revenue, which was finally converted to \BTC and transferred to Bitcoin.}
\label{fig:attack_example}
\end{figure*}

\begin{itemize} [leftmargin=*]
    \setlength\itemsep{0.2mm}
    \item The attacker first withdrew $10$ \ETH from the on-chain mixer, \TC, to hide its identity while use the \ETH to pay the attacking transaction fees afterward (Step~\protect\bcircled{1}).
    
    \item The attacker flash loaned $18{,}308{,}555.417594$ \USDT and $50{,}000{,}000$ \USDC from Uniswap (Step~\protect\bcircled{2}).
    
    \item Next, the attacker swapped $17{,}222{,}012.640506$ \USDT for $17{,}216{,}703.208672$ \USDC in Curve.fi Y pool. This led to the price decrease of \fUSDC in Harvest.Finance USDC vault that uses the Curve.fi Y pool as its price oracle (Step~\protect\bcircled{3}). 
    
    \item The attacker deposited $49{,}977{,}468.555526$ \USDC into the Harvest USDC vault, receiving $51{,}456{,}280.788906$ \fUSDC at $price_{\USDC\rightarrow\fUSDC} = 0.97126080216$ (Step~\protect\bcircled{4}).
    
    \item The attacker exchanged $17{,}239{,}234.653146$ \USDC into $17{,}230{,}747.185604$ \USDT via the Curve.fi Y pool. This exchange caused the price increase of \fUSDC in Harvest.Finance USDC vault (Step~\protect\bcircled{5}).
    
    \item The attacker withdrew from the Harvest USDC vault by trading all \fUSDC shares to $50{,}596{,}877.367825$ \USDC, at $price_{\USDC\rightarrow\fUSDC} = 0.98329837664$ (Step~\protect\bcircled{6}). 
    
    \item The attacker also repaid the flash loans on Uniswap (Step~\protect\bcircled{7}). The net profit of the (not accounting for the flash loan fees) was $619{,}408.812299$ \USDC.
    
    \item All Steps~\protect\bcircled{2} -~\protect\bcircled{7} occurred within a single transaction. The attacker executed $17$ attack transactions against Harvest.Finance USDC vault within $4$ minutes, and $13$ transactions targeting the USDT vault within another $3$ minutes. The attacking transactions finally yielded $13$m \USDC and $11$m \USDT revenue.

    \item To hide the identity, the attack revenue was transferred to the address \texttt{0x381...19B} (Step~\protect\bcircled{8}), which will be then converted to \WBTC, \renBTC and \BTC via Uniswap, Curve.fi, and finally the cross-chain platform, Ren Bridge.
\end{itemize}

The case study shows that sophisticated \DeFi actors can combine multiple techniques for market misconduct. By leveraging techniques such as price oracle manipulation, transaction ordering manipulation, and smart contract exploitation, these actors orchestrate complicated strategies to exploit participants within the \DeFi market. This underscores the need for effective detection and regulatory oversight of such misconduct within the \DeFi market.


\section{Policy implication} \label{sec: policy}

This section explores \textbf{RQ4: What are the regulatory challenges and policy implications associated with market misconduct in the \DeFi ecosystem?} Addressing the policy implications of the emerging \DeFi market misconduct is crucial for tackling the unique challenges it presents. We analyze the limitations of existing regulatory frameworks and identify potential avenues for innovative regulatory solutions.

\subsection{MEV Legality}\label{sec:mev}

To understand the policy implications of market misconduct in \DeFi, we first examine the legality of MEV extractions, categorizing MEV strategies as our analytical foundation. As discussed in Section~\ref{sec: misconduct-defi}, much of the misconduct in the \DeFi market is closely linked to MEV extraction. We then assess how current regulations may apply to \MEV extraction, identifying possible legal gaps. Lastly, we propose responses for regulators to address the associated challenges.

\vspace{+1mm}
\textbf{Toxic and non-toxic MEV.} The classification of MEV extraction into toxic and non-toxic categories is based on its impact on the affected transaction. Toxic MEV refers to strategies that exploit vulnerabilities or manipulate market conditions to harm other participants, such as front-running, sandwich attacks, and other manipulative tactics that create unfair advantages. In contrast, non-toxic MEV includes strategies such as legitimate arbitrage or profit-seeking that leverage decentralized protocols without directly causing harm or manipulating the market. Non-toxic MEV aligns with market efficiency and avoids exploiting participants. As noted by~\cite{qin2022quantifying}, MEV extraction techniques include arbitrage, liquidation, sandwich attacks, and generalized front-running. Among these, sandwich attacks are considered toxic due to their detrimental impact on victim transactions, causing increased price slippage and financial losses.

\vspace{+1mm}
\textbf{Monarch, mafia, and moloch MEV.} MEV is not a single phenomenon but encompasses various forms. \cite{sun2022MEV} classifies MEV into three types: monarch, mafia, and moloch MEV, highlighting the motivations, strategies, and implications of MEV extraction.
Monarch MEV refers to the commonly understood form of MEV, where actors with control over transaction ordering within a block exploit this power to prioritize their transactions or manipulate the market for personal gain.
Mafia MEV arises when an actor uses asymmetric knowledge of another's private information, as in sandwiching \POFs, where adversaries exploit their superior understanding of market dynamics or trading strategies.
Moloch MEV stems from inefficiencies in transaction ordering, such as time-based methods. While chronological ordering appears fair, it fails to account for transaction complexity. Transactions requiring more computation or smart contract interactions may take longer to process, creating inequities even when submitted simultaneously.

The question of whether to impose a comprehensive ban on certain MEV activities remains a subject of debate among industry practitioners and researchers. Classifying MEV as toxic or non-toxic based solely on its impact on individual transactions provides limited insight into its broader market effects. While such classifications help identify immediate harms to participants, they may overlook the cumulative impact on market fairness and efficiency. For example, certain MEV strategies, even when considered non-toxic in isolation, could collectively contribute to market distortions, increased transaction costs, or reduced market efficiency. This underscores the importance of understanding the overall impact of MEV extraction.

On market fairness, \cite{barczentewicz2023mev} argues that certain MEV activities, such as sandwich attacks, may not be inherently unfair, as participants implicitly consent to the transaction price, similar to limit orders on traditional exchanges. \cite{fox2019new} and \cite{barczentewicz2023mev} further emphasize the challenges in defining fairness, while \cite{fletcher2018legitimate} suggests that the claims of open-market manipulation should focus on harm to market efficiency rather than fairness. This perspective underscores the importance of prioritizing market efficiency when assessing MEV.

Regarding market efficiency, MEV has both positive and negative effects. On the positive side, activities such as arbitrage reduce price discrepancies across platforms, enhancing liquidity and market function~(\citealp{qin2022quantifying}). Additionally, MEV redistribution can incentivize validators to engage in block construction, potentially improving network security~(\citealp{chitra2022improving}). On the negative side, MEV can harm retail users through worse execution prices and higher slippage, as noted in Section~\ref{sec:defi_unique}, discouraging participation. Excessive MEV extraction can also incentivize validators to reorganize blocks, destabilizing blockchain consensus. While MEV can promote efficiency in certain cases, its broader impacts—both beneficial and harmful—remain widely debated~(\citealp{barczentewicz20blockchain}). Balancing these effects is crucial to address MEV’s complex implications for market dynamics and blockchain stability.

Assessing the individual impact of MEV differs from evaluating its broader social impact, yet existing studies~(e.g., \citealp{daian2020flash,zust2021analyzing,qin2021attacking,xiong2023demystifying}) primarily focus on individual effects, often neglecting the social dimension. As \cite{chitra2022improving} suggests, MEV activities such as sandwich attacks may harm individual traders while benefiting block validators. Similarly, JIT liquidity attacks can improve execution prices for traders involved but negatively affect existing liquidity providers in the pool. Thus, focusing solely on individual welfare fails to capture the full scope of MEV’s impact, as negative individual experiences may be offset by broader market efficiency gains. 
The unclear social impact of MEV leads to the ongoing debate about its legality, with some researchers~(e.g., \citealp{barczentewicz20blockchain}) arguing that certain MEV activities could be manipulative, potentially violating rules such as \CFTC \href{https://www.law.cornell.edu/cfr/text/17/180.1}{Rule 180.1} or \SEC \href{https://www.law.cornell.edu/cfr/text/17/240.10b-5}{Rule 10b-5}. The central question is whether MEV constitutes unacceptable market manipulations or legitimate strategies. \cite{fletcher2018legitimate} states that if a strategy enhances or does not harm market efficiency, it should not be classified as manipulation. From this perspective, MEV’s legality depends on its net impact on market efficiency. However, the limited understanding of MEV’s overall impact presents challenges for policymakers. Therefore, further academic research is crucial to evaluate MEV’s net impact on the market and social welfare, provide insights into its consequences, and guide more informed regulatory decisions.

\subsection{Regulatory Challenges} \label{sec:challenges}

Advancements in financial technology pose significant regulatory challenges, as existing laws often fall short of addressing their impact. The rise of \DeFi market misconduct further exacerbates these challenges, presenting regulators with complex issues requiring urgent attention.

\textbf{Blockchain pseudonymity and privacy.} Unlike \TradFi, where personal information is often collected and verified through the \KYC processes, blockchain transactions in \DeFi are often pseudonymous, with participants identified by their wallet addresses rather than personal information. While this nature of \DeFi offers benefits such as increased privacy and reduced reliance on centralized authorities, it also poses challenges for regulatory oversight. The absence of personal information linked to on-chain addresses makes it difficult for regulators to trace and attribute transactions to specific individuals or entities. This limitation hampers their ability to detect and prevent market misconduct. Without accurate identification, regulators face challenges in assessing risk, conducting investigations, and enforcing compliance with the \AML requirement. Addressing this challenge requires finding a balance between preserving user privacy and ensuring regulatory compliance. Regulators are exploring various approaches to tackle the pseudonymity challenge in \DeFi. This involves adopting enhanced transaction monitoring tools, utilizing detailed blockchain analytics for oversight, and exploring decentralized identity protocols that balance identity verification with user privacy.

\textbf{Blockchain decentralization.} The decentralization of blockchain technology, particularly in emerging markets such as \DeFi, presents unique challenges for regulators to address misconduct behaviors. One of the core issues is that blockchain decentralization makes it difficult to determine who should be held liable for such misconduct, even if the adversary's identity can be detected. In \TradFi, regulatory authorities often have a clear framework to identify and hold accountable the parties responsible for misconduct. This is achieved through various mechanisms, including the identification and verification of individuals or entities, centralized control over transactions, and the ability to enforce regulatory measures on centralized intermediaries. However, in decentralized systems like \DeFi, these mechanisms are fundamentally different. Participants in \DeFi transactions are identified through their pseudonymous addresses rather than personal information. While the adversary's identity may be detectable through sophisticated forensic techniques~(\citealp{biryukov2019deanonymization}), determining legal liability is difficult. When misconduct occurs in \DeFi, it becomes difficult to attribute responsibility to specific entities. The absence of centralized intermediaries or authorities makes it challenging for regulators to enforce accountability against malicious actors. Blockchain decentralization often leads to a diffusion of responsibility, whereby numerous participants could have been involved in a particular instance of misconduct to varying degrees.

\textbf{Blockchain immutability.} The immutability of blockchain, a fundamental characteristic that ensures the integrity and security of transactions, presents a significant challenge for regulators. In the \TradFi system, regulators and authorities have the ability to reverse fraudulent transactions or recover stolen funds through various mechanisms, such as freezing accounts, conducting investigations, and enforcing legal actions. Nevertheless, within the decentralized and immutable framework of blockchain, once a transaction is included in a block, it becomes impossible to alter or erase. When market misconduct occurs in \DeFi, the stolen funds are often irreversibly transferred to the adversary's address. Even if regulators identify the market misconduct and the address to which the stolen funds were transferred, the decentralized and immutable characteristics of the blockchain make it extremely challenging to retrieve the funds and return them to the victims. The stolen assets become permanently locked within the blockchain, beyond the reach of traditional regulatory interventions. This lack of transaction reversibility presents a significant obstacle for regulators seeking to provide restitution to affected individuals or take action against malicious actors. Consequently, victims may permanently lose their assets without any feasible recourse.

\textbf{Blockchain borderlessness.} In \TradFi, regulatory authorities operate within defined jurisdictions, enforcing rules based on established frameworks. In contrast, blockchain operates globally, enabling \DeFi participants from any location to transact without geographical constraints. This borderlessness poses challenges for regulators, as \DeFi misconduct often involves participants across multiple jurisdictions, complicating the application of regulatory frameworks.  Each country has its own legal and regulatory systems, which may have varying definitions, interpretations, and enforcement mechanisms. Harmonizing regulatory approaches and aligning enforcement efforts across multiple jurisdictions can be a time-consuming and challenging process. First, regulators must navigate the complexities of international law and differences in regulatory philosophies. The lack of a global consensus on jurisdictional standards can lead to regulatory arbitrage, where participants exploit regulatory differences by engaging in \DeFi activities from jurisdictions with less stringent regulations. Additionally, different legal and cultural contexts across countries can also impact the perception and definition of misconduct. What may be considered misconduct in one jurisdiction may not be viewed the same way in another. This divergence further complicates the establishment of a global standard for jurisdiction in regulating emerging market misconduct in \DeFi.

\textbf{Regulation vs. innovation.} Bring \DeFi into the regulatory perimeter presents a challenge in finding the balance between regulation and innovation. Excessive regulation can stifle innovation, while inadequate regulation can leave market participants vulnerable to misconduct. As mentioned in Section~\ref{sec:mixer}, the on-chain mixer service \TC was sanctioned by the U.S. \OFAC on Aug $8$,~$2022$. According to the \OFAC, \TC has been implicated in the laundering of cryptocurrency assets valued at over $7$b \USD since its establishment in $2019$\footnote{\href{https://home.treasury.gov/news/press-releases/jy0916}{U.S. Treasury Sanctions Notorious Virtual Currency Mixer Tornado Cash}.}. On one hand, regulation is necessary to protect market participants. Regulatory frameworks help establish rules to prevent fraudulent activities, market manipulation, and other forms of misconduct. On the other hand, the sanction on \TC may raise concerns among privacy-conscious users. While the sanction targets the illicit activities associated with \TC, it can potentially impact legitimate users who rely on the mixer for legitimate purposes, such as protecting their privacy. Hence, finding the right balance is crucial in promoting the growth of innovation while safeguarding market participants.


\subsection{Early Regulatory Responses} \label{sec:early_resp}

This section first presents the varying regulatory frameworks for cryptocurrency in major global economies, while also analyzing their approaches to \DeFi regulation.

\textbf{United States.} In the U.S., various authorities, including the \SEC, \CFTC, and \FinCEN, exercise jurisdiction over different aspects of cryptocurrencies. A prerequisite for the jurisdiction of the \SEC is that the activity in question involves securities\footnote{Section 2(a)(1) of the U.S. Securities Act of 1933 and Section 3(a)(10) of the \SEA 1934 each list a variety of assets that are considered securities.}. In the context of \DeFi, the \SEC can potentially exercise regulatory authority over specific tokens and platforms, contingent upon their classification as securities under the Howey Test\footnote{The Howey Test, derived from a landmark U.S. Supreme Court case, is used to determine whether a transaction would be considered a security and subject to disclosure and registration requirements. See \href{https://www.sec.gov/corpfin/framework-investment-contract-analysis-digital-assets}{Framework for ``Investment Contract'' Analysis of Digital Assets} for more details.}. Should these tokens be deemed securities, they would be subject to regulatory obligations such as registration, disclosure, and other compliance requirements. For instance, \SEC filed a lawsuit against Ripple Labs in $2020$, alleging that its \XRP token was an unregistered security\footnote{See \href{{https://www.sec.gov/news/press-release/2020-338}}{SEC Charges Ripple and Two Executives with Conducting $1.3$b Unregistered Securities Offering}.}.

\CFTC, on the other hand, is responsible for regulating the trading of commodity derivatives. \DeFi platforms that offer products based on commodities may come under the jurisdiction of the \CFTC. The \CEA enables the \CFTC to regulate the trading of commodities and futures contracts and supervise market participants. The \CEA also prohibits market manipulation practices in commodity trading and provides the \CFTC with enforcement powers to prosecute violations. During a keynote address presented at the Brookings Institution's webcast, \CFTC chair Rostin Behnam has expressed concerns regarding the involvement of digital assets and \DeFi in commodity derivatives activity\footnote{See \href{https://www.cftc.gov/PressRoom/SpeechesTestimony/opabehnam24}{Keynote Address of Chairman Rostin Behnam at the Brookings Institution Webcast on The Future of Crypto Regulation}.}. Indeed, he has expressed a strong attitude towards regulating the \DeFi market: ``\DEXs and other aspects of the \DeFi world can be regulated in the U.S., even though it is just code''\footnote{See \href{https://www.bloomberg.com/news/articles/2023-05-18/cftc-chair-rostin-behnam-discusses-crypto-regulation-on-odd-lots}{CFTC Chair Rostin Behnam on the Fight to Regulate Crypto}.}. At the Senate Agricultural hearing of $2023$, Rostin Behnam also reaffirmed that stablecoins and \ETH were commodities\footnote{See \href{https://cointelegraph.com/news/stablecoins-and-ether-are-going-to-be-commodities-reaffirms-cftc-chair}{Stablecoins and Ether are going to be commodities}.}.  

\FinCEN, the agency responsible for enforcing AML and \CTF regulations, has provided guidance indicating that certain activities in the DeFi space may be considered as Money Services Businesses (MSBs)\footnote{See \href{https://www.fincen.gov/resources/statutes-regulations/guidance/application-fincens-regulations-persons-administering}{Application of \FinCEN's Regulations to Persons Administering, Exchanging, or Using Virtual Currencies}}. Consequently, these \DeFi platforms should conform to the BSA's provisions and establish robust AML/CTF procedures. As discussed in Section~\ref{sec:intro}, the recent ``Illicit Finance Risk Assessment of Decentralized Finance''\footnote{\href{https://home.treasury.gov/system/files/136/DeFi-Risk-Full-Review.pdf}{Illicit Finance Risk Assessment of Decentralized Finance}} released by the U.S. Department of the Treasury expresses concerns and regulatory expectations with \DeFi activities. It highlights that ``the most significant current illicit finance risk in this domain is from \DeFi services that are not compliant with existing \AML/\CTF obligations''.

\textbf{United Kingdom.} In Jun $2022$, the \HMT preleased the results of its $2021$ consultation on the proposed changes to the UK \href{https://www.gov.uk/government/consultations/amendments-to-the-money-laundering-terrorist-financing-and-transfer-of-funds-information-on-the-payer-regulations-2017-statutory-instrument-2022}{Money Laundering Regulations}, along with the suggested actions in response. The \HMT proposed to collect information on entities involved in unhosted wallet transfers, where the corresponding transactions pose an elevated risk of illicit finance. This proposal necessitates the regulation of crypto service providers to fulfill \AML and \CTF requirements. 

In Feb $2023$, the \HMT released its consultation on the future financial services regime for crypto-assets, including a proposal for a crypto-asset market abuse regime\footnote{\href{https://www.gov.uk/government/consultations/future-financial-services-regulatory-regime-for-cryptoassets}{Future financial services regulatory regime for cryptoassets}.}. Similar to the U.S., the UK government decides that crypto-assets should be regulated within the existing \FSMA framework. According to this consultation, crypto-asset trading venues are required to establish robust systems to prevent and detect any instances of market abuse. Additionally, these venues are expected to carry out investigations and impose sanctions for market abuse, under the supervision of the \FCA. Moreover, all regulated firms engaged in crypto-asset activities are required to disclose inside information and maintain insider lists. However, the \HMT has not yet introduced definitive regulations specifically targeting the \DeFi market. Instead, the regulators are actively seeking evidence to better understand the regulatory landscape and to formulate effective strategies.

\textbf{European Union.} 
Since the release of the $2018$ \href{https://eur-lex.europa.eu/legal-content/EN/TXT/HTML/?uri=CELEX:52018DC0109&from=EN}{Fintech Action Plan}, the EU Commission has been analyzing the potential and challenges presented by crypto-assets. In Sep $2020$, the Commission introduced \MiCA regulation, as part of the \href{https://finance.ec.europa.eu/publications/digital-finance-package_en}{Digital Finance Package}. Unlike the U.S. and the UK, which integrate cryptocurrency regulation into existing frameworks, the EU has introduced the \MiCA regulation as a bespoke regime for crypto-assets.
The primary goal of \MiCA is to harmonize the regulation of crypto-assets and distributed ledger technology (DLT) across the EU. The \MiCA provides regulatory definitions for crypto-assets, including utility tokens, asset-referenced tokens, e-money tokens, and other categories of crypto-assets. 
The \MiCA framework establishes two key regulatory regimes: one for Crypto-Asset Service Providers (CASPs), outlining operational, licensing, and conduct requirements, and another for issuers of crypto-assets, focusing on transparency, disclosure obligations, and consumer protection standards.
Moreover, \MiCA establishes a comprehensive framework for preventing market abuse, although this framework may not be applied directly to regulate the \DeFi market.

As discussed in Section~\ref{sec: existing_defi}, the \DeFi exclusion from the MiCA is another topic of continuing discussion. The MiCA's specification that ``Where crypto-asset services as defined in this Regulation are provided in a fully decentralized manner without any intermediary they do not fall within the scope of this Regulation'' has raised significant concerns. Consequently, \MiCA's prohibition of market manipulation, insider dealing, and unlawful disclosure of inside information may not be directly applied to regulate \DeFi market misconduct. In fact, the Association for Financial Markets in Europe (AFME) has raised concerns about \DeFi's exclusion from the \MiCA, arguing that doing so might compromise the efficiency of the EU's regulatory frameworks\footnote{See \href{https://www.afme.eu/Portals/0/DispatchFeaturedImages/AFME\%20DeFi\%20Whitepaper.pdf}{AFME, Principles for building a robust digital economy}.}. 
Excluding decentralized activities from regulation may threaten financial stability and create a fertile ground for potential ``regulatory arbitrage'' opportunities that malicious actors could exploit. 
Indeed, the \FSB recognizes the importance of proactively assessing the financial vulnerabilities of the \DeFi ecosystem. In their recent report\footnote{See \href{https://www.fsb.org/2023/02/the-financial-stability-risks-of-decentralised-finance/}{FBS, The Financial Stability Risks of Decentralised Finance}.}, they assess the financial vulnerabilities of \DeFi and emphasize the significance of including \DeFi in the regular monitoring of the broader crypto-asset markets. In summary, \MiCA's exclusion of DeFi highlights the significance of assessing the level of decentralization in \DeFi protocols and determining the level of decentralization that holds relevance for regulation.

\textbf{China.} China has implemented strict regulations on cryptocurrency over the past decade. In 2013, the People's Bank of China (PBOC) and several other agencies issued a joint notice prohibiting financial institutions from handling Bitcoin transactions\footnote{See \href{https://www.gov.cn/gzdt/2013-12/05/content_2542751.htm}{Notice on Preventing Bitcoin Risks}.}, marking the first official regulatory action against cryptocurrencies. In 2017, the government strengthened its measures by banning initial coin offerings (ICOs) and requiring the closure of domestic cryptocurrency exchanges\footnote{See \href{https://www.gov.cn/xinwen/2017-09/04/content_5222657.htm}{Announcement on Preventing Financial Risks from ICOs}.}, effectively halting centralized trading activities within the country. In 2021, regulatory pressure intensified as the government declared that all cryptocurrency-related transactions were illegal, including trading and mining. 

As for \DeFi, while China has not implemented direct regulations specifically targeting \DeFi, its general ban on cryptocurrency-related activities effectively restricts \DeFi operations. Additionally, the China Financial Stability Report (2023)\footnote{\href{http://www.pbc.gov.cn/goutongjiaoliu/113456/113469/5177895/2024071920310922075.pdf}{China Financial Stability Report (2023)}.} highlights significant concerns regarding \DeFi, including its susceptibility to centralized control by dominant participants, its potential to facilitate illicit activities due to transactional anonymity, and the regulatory complexities arising from its cross-border nature.

\textbf{Janpan.} Japan regulates cryptocurrency through a dual classification system: crypto assets under the \href{https://www.japaneselawtranslation.go.jp/en/laws/view/3965/en}{\PSA} and security tokens under the \FIEA. Crypto assets, used as payment or value transfer mechanisms, are regulated by the \FSA with a focus on exchange registration, AML compliance, and asset protection measures. Security tokens, representing investment rights such as equity or debt, are treated as securities and subject to strict disclosure requirements, insider trading prohibitions, and anti-manipulation rules under the \FIEA.  Similar to the regulatory approaches in the UK and the U.S., Japan has not introduced specific regulations for \DeFi. Instead, it relies on existing laws, such as the \PSA and \FIEA, to oversee \DeFi activities that may intersect with regulated financial services, focusing on areas such as \AML and investor protection.

\smallskip
Overall, our analysis of the regulatory frameworks of major global economies reveals distinct approaches: the UK, the U.S., and Japan regulate the crypto space through existing financial regulatory frameworks, the EU has established a bespoke regime for crypto regulation, and China enforces a general ban on all crypto-related activities. Several key findings emerged from our analysis when examining the initial regulatory response to the DeFi market. 
\begin{itemize}[leftmargin=*]
    \setlength\itemsep{0.1mm}
    \item \textbf{Diverse regulatory approaches}: Different jurisdictions have taken diverse methods to regulate crypto activities. Some adopt existing regulations, while others seek to establish bespoke regulatory regimes.
    
    \item \textbf{Absence of DeFi-specific regulations}: While DeFi activities may indirectly be covered by existing regulations, there is an absence of direct regulations specifically tailored to address the unique challenges in the \DeFi space.
    
    \item \textbf{Undefined metrics for decentralization level}: While the level of decentralization is a crucial metric for determining the regulatory implications and risk profiles of DeFi protocols, there is currently no consensus on how to assess DeFi protocols' decentralization levels and how to determine its regulatory significance.
    \item \textbf{Challenges in regulating DeFi market misconduct}: The DeFi market lacks a clear legal definition and taxonomy for misconduct activities. Moreover, enforcing accountability when penalizing misconduct poses significant challenges.
\end{itemize}

\subsection{Recommendations}

\subsubsection{Regulatory Efforts}

While the decentralized and pseudonymous nature of \DeFi challenges user-level regulation, enforcing rules at the protocol level remains feasible. Unlike \TradFi, where regulators oversee centralized entities and enforce compliance directly on users, \DeFi operates through smart contract protocols that govern \DApps. Regulating these protocols, rather than individual users, can potentially establish responsible behavior and mitigate market misconduct risks.

\vspace{+1mm}
\textbf{AML/KYC + Money Tracing.} In \TradFi, \AML/\KYC requirement is crucial to prevent money laundering, terrorist financing, and other illicit activities. Similarly, \CeFi platforms, such as \CEXs, are also subject to various \AML obligations. The U.S. \href{https://www.fincen.gov/resources/statutes-and-regulations/bank-secrecy-act}{Bank Secrecy Act (BSA)} imposes a range of obligations on financial institutions. According to the BSA, any \DeFi service that functions as a financial institution, regardless of whether it is centralized or decentralized, should be required to comply with \AML obligations. However, many existing \DeFi services covered by the BSA fail to meet AML requirements\footnote{\href{https://home.treasury.gov/system/files/136/DeFi-Risk-Full-Review.pdf}{Illicit Finance Risk Assessment of Decentralized Finance}}. Indeed, incorporating \AML/\KYC protocols into DeFi presents significant complexity. These protocols require the verification and collection of user data. Traditional \AML/\KYC solutions depend on centralized entities for user identity verification, creating obstacles when adapting to \DeFi. Moreover, developing decentralized identity verification mechanisms that uphold data integrity while ensuring compliance remains an ongoing challenge. 

\begin{figure}[tbh]
\centering
\includegraphics[width=\columnwidth]{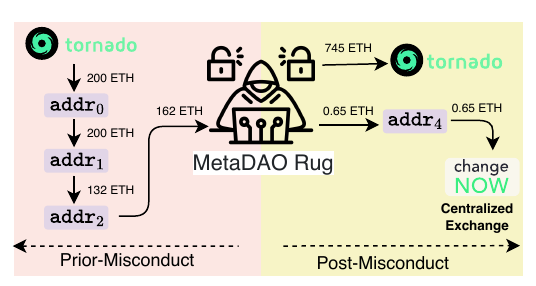}
\caption{Motiviting example of regulating DeFi market misconduct by tracing the source and destination of funds. The adversary first withdrew initial funds from the on-chain mixer, \TC, via three intermediary addresses. After the rug pull, the adversary deposited the revenue of $745$ \ETH primarily into TC, with the intention of concealing its identity. In the meanwhile, the adversary also transferred a smaller amount of $0.65$ \ETH to a CEX, located at \href{hangenow.io/}{https://changenow.io/}. This transfer creates a potential traceability avenue for the regulator if the CEX has implemented AML/KYC requirements.
}
\label{fig:regulation_example}
\end{figure}

The challenges of implementing \AML/\KYC measures for \DeFi protocols do not suggest that regulating market misconduct in the \DeFi space is an unattainable goal. Figure~\ref{fig:regulation_example} provides a motivating example of how \DeFi market misconduct can be potentially regulated by tracing the source and destination of funds. In the MetaDAO rug pull example\footnote{\url{https://twitter.com/PeckShieldAlert/status/1475434691939520523}}, the adversary first withdrew initial funds from the on-chain mixer, \TC, via three intermediary addresses. After the rug pull, the adversary deposited the illicitly obtained revenue of $745$ \ETH primarily into the on-chain mixer service, TC, with the intention of concealing its identity. In the meanwhile, the adversary also transferred a smaller amount of $0.65$ \ETH to a CEX named \href{https://changenow.io/}{NOW exchange}. This transfer creates a potential traceability avenue for the regulator if the \CEX has implemented \AML/\KYC.

This example underscores the importance of tracing adversarial funds within the \DeFi ecosystem. Future research should focus on developing advanced money-tracing tools to help regulators track fund flows, detect misconduct, and support enforcement actions. 
Moreover, regulators can also incentivize financial institutions to adopt robust \AML/\KYC measures. If money-tracing tools reveal that adversarial funds originated from a \CEX, this strengthens the case for regulatory action and holds the CEX accountable for enabling illicit activities. Such measures encourage platforms to prioritize compliance, fostering a regulatory environment that promotes accountability and reduces misconduct.

\vspace{+1mm}
\textbf{Smart Contract Audit Framework.} Requiring \DeFi protocol development teams to conduct security audits of their smart contracts can mitigate misconduct stemming from code vulnerabilities. Regulators can collaborate with the industry to establish a security auditing framework that prioritizes key areas such as smart contracts, \DApps, tokenomics, and protocol architecture. The framework could include:
\emph{(i)} Reviewing the smart contract codebase to identify vulnerabilities, including coding errors, security loopholes, and adherence to secure coding standards.
\emph{(ii)} Conducting vulnerability testing through automated tools and manual inspection to uncover both known and novel weaknesses.
\emph{(iii)} Assessing tokenomics to detect risks such as front-running, flash loan attacks, or economic manipulation, ensuring the economic model aligns with the protocol’s objectives.
\emph{(iv)} Evaluating governance structures and processes to ensure resilience against governance attacks and manipulation.
\emph{(v)} Documenting audit findings with actionable recommendations and clear risk assessments for stakeholders.
Besides, encouraging \DeFi platforms to implement continuous monitoring systems and establish bug bounty programs can further enhance security by detecting emerging threats and incentivizing external researchers to report vulnerabilities.

\vspace{+1mm}
\textbf{Market Protection Mechanisms.} Market protection mechanisms in \TradFi, such as circuit breakers, can be adapted for the \DeFi ecosystem to mitigate misconduct. Circuit breakers temporarily halt or limit trading during extreme market volatility to prevent cascading disruptions~(\citealp{wang2022asymmetric}). In \DeFi, these mechanisms can be applied as follows:
\emph{(i)} Volatility control: Circuit breakers can activate based on predefined volatility thresholds, pausing or limiting specific activities such as trading or lending to stabilize the market and prevent manipulation.
\emph{(ii)} Smart contract interventions: In cases of misconduct or suspicious activity, circuit breakers can freeze or disable specific smart contract functions to prevent further losses.
\emph{(iii)} Gradual reactivation: After activation, circuit breakers can enable a controlled and phased resumption of \DeFi activities to ensure a smooth recovery.
To ensure effectiveness, circuit breakers should be transparently pre-defined within public smart contracts. Developers must carefully design these mechanisms to prevent unintended vulnerabilities, particularly given the composability of \DeFi protocols.

\vspace{+1mm}
\textbf{Generalized Accounting Framework.} In \TradFi, accounting frameworks such as Generally Accepted Accounting Principles (GAAP) and International Financial Reporting Standards (IFRS) provide clear guidelines for recording financial transactions, preparing statements, and ensuring transparency and accuracy. Institutions maintain ledgers to track assets, liabilities, income, and expenses, with external audits verifying compliance with these standards.

Efforts to establish accounting frameworks for digital assets have been made, but no standardized model exists for \DeFi protocols. For instance, the Financial Accounting Standards Board (FASB) proposed an update in March 2023 to improve accounting and disclosure for crypto-assets\footnote{See \href{https://fasb.org/page/getarticle?uid=fasb_Media_Advisory_03-23-23}{FASB seeks public comments on proposed improvements to the accounting for and disclosure of certain crypto assets.}}, while the International Swaps and Derivatives Association (ISDA) explored accounting implications for digital assets\footnote{See \href{https://www.isda.org/2022/05/10/accounting-for-digital-assets-key-considerations/}{Accounting for Digital Assets: Key Considerations}}.

Blockchain-based \DeFi protocols offer transparent, immutable ledgers, making it feasible to develop standardized accounting frameworks. \cite{lommers2022dao} proposed a framework for \DAOs, incorporating principles of double-entry bookkeeping adapted to \DeFi. This framework captures diverse financial activities like staking, lending, token minting/burning, and airdrops, demonstrating its potential for comprehensive financial documentation within \DeFi platforms. Such frameworks could also serve regulatory purposes, enabling the detection of unusual patterns indicative of market manipulation or fraud, such as excessive fund inflows or pyramid-like schemes. Financial ratios and risk assessments derived from these systems could further support risk management in \DeFi. However, implementing a generalized framework for \DeFi presents challenges due to the diversity and complexity of financial instruments in the ecosystem. Regulators should carefully adapt the framework to account for the unique characteristics of various \DeFi instruments.

\vspace{+1mm}
\textbf{AI Integration.} The advancement of AI technology provides regulators with an effective tool to detect market misconduct. In \TradFi, AI has been widely adopted to detect trade-based market manipulation.
For example, \cite{wang2019enhancing} proposed a recurrent neural network (RNN)-based ensemble learning framework, a supervised learning approach, that effectively detects stock price manipulation activities by combining trade-based features derived from trading records with characteristic features of listed companies. As for unsupervised learning, \cite{kamps2018moon} proposed a novel hybrid anomaly detection method for pump and dump detection based on distance and density metrics. 

AI techniques have also demonstrated potential in detecting DeFi market misconduct. For instance, studies such as \cite{hu2023sequence,chadalapaka2022crypto,nghiem2021detecting,bello2023lld} employed neural networks, while \cite{la2023doge} applied random forest and Adaboost to identify cryptocurrency pump-and-dump schemes. Although these methodologies were developed in \CeFi contexts, they could be adapted for \DeFi, considering its unique characteristics. Additionally, several studies have specifically analyzed wash trading within \DeFi using on-chain data. For example, \cite{cui2023wteye} proposed algorithms to detect wash trading in ERC20 token transactions, finding that over 15\% of activity involves wash trading. Similarly, \cite{victor2019cryptocurrency} used a graph model to examine wash trading on \DEXs like IDEX and Etherdelta, revealing that at least 30\% of their activities are likely to be wash trades.

\vspace{+1mm}
\textbf{Quantum Integration.} Quantum computing has significant potential to enhance regulators’ ability to detect misconduct in DeFi markets, though its current applications are largely theoretical. The technology’s capacity to solve complex optimization problems and analyze large datasets~(\citealp{acharjya2016survey}) could enable regulators to process \DeFi transaction patterns and smart contract activities more effectively than classical computing methods. For instance, quantum algorithms could help identify manipulations such as front-running, sandwich attacks, and oracle manipulation by detecting anomalies within blockchain networks.

However, the practical application of quantum computing in DeFi regulation is limited by several challenges. Quantum technology is still in its early stages, with only a few operational quantum computers capable of performing these tasks. Additionally, applying quantum algorithms to DeFi’s decentralized and pseudonymous framework would require substantial advancements in quantum machine learning and their integration with blockchain systems. Overall, although quantum computing is not yet a viable tool for detecting DeFi market misconduct, its potential to address the scale and complexity of decentralized ecosystems suggests it may become an essential resource as technology advances.

\vspace{+1mm}
\textbf{Cross-Country Coordination.} 
As discussed in Section~\ref{sec:challenges}, \DeFi's borderless operations challenge regulators limited by jurisdictional boundaries and frameworks. Harmonizing regulations across jurisdictions is complex due to varying national approaches to \DeFi market misconduct. Potential solutions include:
\emph{(i)} International coordination: Encouraging collaboration among regulatory bodies, such as through the \FSB or international standard-setting organizations, to exchange information, best practices, and regulatory strategies.
\emph{(ii)} Cross-jurisdictional task forces: Establishing task forces or working groups, such as the Blockchain and Virtual Currency Working Group (WG), to address market misconduct and align regulatory efforts.
While full harmonization may be impractical given differing legal frameworks, collaboration and information sharing can promote a more consistent regulatory environment.

\vspace{+1mm}
\subsubsection{Academic Efforts}
Future academic efforts should focus on several areas in light of the ongoing challenges.

\begin{itemize}[leftmargin=*]
    \setlength\itemsep{0.1mm}
    \item \textbf{Assessing the net impact of \DeFi market misconduct}. As outlined in Section~\ref{sec:mev}, it is important for researchers to investigate the overall impact of misconduct behaviors on market efficiency and social welfare. This requires expanding the focus from the individual impacts to the analysis of market-wide implications. This method enables the discovery of critical insights into the legality of such behaviors and the effects on the wider DeFi ecosystem, thereby assisting regulators in developing effective regulatory frameworks and making informed decisions. 
    
    \item \textbf{Building reliable metric to measure the level of decentralization}. Section~\ref{sec:early_resp} shows that DeFi's exclusion from the \MiCA regulation raised significant concerns, as the degree of decentralization directly influences the regulatory implications and risk assessment. With reliable decentralization metrics, regulators can adopt a tailored approach that matches the unique characteristics and risks associated with each protocol. Decentralized protocols may require lighter regulatory oversight, recognizing their self-regulating nature, while more centralized ones may demand closer supervision to prevent fraudulent activities.
    \item \textbf{Developing robust risk assessment tools for DeFi protocols}. DeFi protocols often interact with multiple components (e.g., smart contracts, oracles, governance mechanisms, etc.), and a vulnerability in any of these components can affect the entire ecosystem. Robust risk assessment tools help identify systemic risks of various \DeFi components, enabling regulators to implement targeted measures to mitigate risks and prevent cascading failures.
\end{itemize}

\section{Discussion}\label{sec: discussion}

Regulatory bodies worldwide are actively exploring approaches to regulate the crypto space and applications of DLT. Initial frameworks, such as the EU’s \MiCA, serve as foundational steps. However, current efforts primarily target \CeFi institutions, while \DeFi remains a subject of concern due to its decentralized and pseudonymous nature. Nonetheless, the significant risks associated with \DeFi underscore the necessity of its integration into regulatory frameworks.

Regulating \DeFi involves multiple dimensions such as risk management, consumer protection, and market integrity. While this study focuses on \DeFi market misconduct, the topic extends well beyond a niche area, given its broader implications for financial stability, market fairness, and consumer trust.  Our analysis demonstrates that emerging forms of misconduct in the nascent \DeFi space allow malicious actors to implement complex strategies to exploit the system, thereby undermining market stability and investor confidence. Effective oversight of market misconduct is crucial to strengthen the credibility of the \DeFi ecosystem and facilitate its integration into global financial systems.

\section{Limitations and Future Research} \label{sec:limitation}
This study analyzes market misconduct under the EU’s \MAR framework, a widely referenced regulatory standard. However, regulatory frameworks differ across jurisdictions, and the evolving nature of the \DeFi market gives rise to new forms of misconduct. Consequently, our proposed taxonomy of \DeFi market misconduct should be considered a foundational framework, requiring adaptation as \DeFi markets and global regulatory landscapes continue to evolve.

Another limitation of this study is the lack of empirical analysis on order-based manipulation. While we provide empirical insights into trade-based and transaction-ordering manipulations in \DeFi, order-based manipulation is excluded due to its reliance on off-chain order matching, which presents challenges in data accessibility. Future work can develop methodologies to obtain and analyze these data, enhancing the understanding of order-based manipulation in \DeFi.

\section{Conclusion} \label{sec: conclusion}

The rise of \DeFi highlights how technological advancement can drive financial innovations. However, these innovations also heighten the risk of misconduct in \DeFi markets. This duality presents both opportunities and challenges for developing a regulatory framework for \DeFi. This study analyzes the potential risks of market misconduct inherent in \DeFi.  We propose a comprehensive taxonomy of \DeFi market misconduct, analyze its characteristics and impact, explore the challenges of developing a tailored \DeFi regulatory framework, and suggest potential approaches for integrating \DeFi into the regulatory perimeter.

Our proposed definition and taxonomy of \DeFi market misconduct offer a structured foundation for understanding behaviors that undermine market efficiency, fairness, and security. The comparison of market misconduct in \DeFi and \TradFi underscores the emergence of novel misconduct forms unique to \DeFi. The case study illustrates how actors can integrate multiple manipulative strategies to exploit \DeFi markets. Furthermore, our analysis of current regulatory frameworks reveals their limitations in addressing \DeFi-specific risks and emphasizes the need for regulatory adaptation to this evolving ecosystem. Finally, we propose potential approaches to integrate \DeFi into the regulatory perimeter. We hope that this study not only advances academic understanding of \DeFi market misconduct but also serves as a resource for policymakers and regulators in developing robust frameworks.

\printcredits

\bibliographystyle{model1-num-names}

\bibliography{ref}

\end{document}